\documentclass{nature_figs}
\usepackage{amsmath}
\usepackage{amssymb}
\usepackage{hyperref}
\usepackage{lineno}
\usepackage{float}
\usepackage{graphicx}%
\usepackage{multirow}%
\usepackage{amsmath,amssymb,amsfonts}%
\usepackage{amsthm}%
\usepackage[title]{appendix}%
\usepackage{xcolor}%
\usepackage[super,comma]{natbib}
\usepackage{etoolbox}
\usepackage{textcomp}%
\usepackage{manyfoot}%
\usepackage{booktabs}%
\usepackage{xspace}
\usepackage{listings}% 
\usepackage{adjustbox}
\usepackage{longtable}
\usepackage{acronym}  % Acronyms
\usepackage{units}
\usepackage[table]{xcolor}

\hypersetup{colorlinks,citecolor=blue,linkcolor=blue,urlcolor=blue}
\acrodef{CC-SN}[CC-SN]{Core Collapse Supernova}
\acrodef{SNR}[SNR]{Supernova Remnants}
\acrodef{XDINS}[XDINS]{X-ray Dim Isolated Neutron Stars}
\acrodef{CCO}[CCO]{Central Compact Objects}
\acrodef{RPP}[RPP]{Rotation-Powered Pulsars}
\acrodef{FRBs}[FRBs]{Fast Radio Burst}
\acrodef{ISM}[ISM]{interstellar medium}

\title{Magnetar fraction in Core-Collapse Supernovae }

\author{Celsa Pardo-Araujo$^{1,2}$,
    Nanda Rea$^{1,2}$,
          Michele Ronchi$^{3}$, and 
           Vanessa Graber$^{4}$
          }

\makeatletter
\let\saved@includegraphics\includegraphics
\AtBeginDocument{\let\includegraphics\saved@includegraphics}
\renewenvironment*{figure}{\@float{figure}}{\end@float}
\makeatother
\begin{document}

\maketitle

% AFFILIATIONS
\begin{affiliations}

  \item Institute of Space Sciences (ICE), CSIC, Campus UAB, Carrer de Can Magrans s/n, E-08193, Barcelona, Spain
  \item Institut d'Estudis Espacials de Catalunya (IEEC), Edifici RDIT, Campus UPC, E-08860 Castelldefels (Barcelona), Spain
%  \item Nordita, KTH Royal Institute of Technology and Stockholm University, Hannes Alfv\'ens v\"ag 12, SE-10691 Stockholm, Sweden

   \item ASTRON, the Netherlands Institute for Radio Astronomy, Oude Hoogeveensedijk 4, 7991 PD Dwingeloo, The Netherlands
   \item Department of Physics, Royal Holloway, University of London, Egham, TW20 0EX, UK
 
\end{affiliations}

%%================================%%
%% Sample for structured abstract %%
%%================================%%

% \abstract{\textbf{Purpose:} The abstract serves both as a general introduction to the topic and as a brief, non-technical summary of the main results and their implications. The abstract must not include subheadings (unless expressly permitted in the journal's Instructions to Authors), equations or citations. 
% 
% \textbf{Methods:} 
% 
% \textbf{Results:} 
% \textbf{Conclusion:} 

% ABSTRACT (200 words)
\begin{abstract}
Magnetars are extreme neutron stars powered by ultra-strong magnetic fields ($\sim10^{14}$\,Gauss) and are compelling engines for some of the most powerful extragalactic transients such as Super Luminous Supernovae, Gamma-Ray Bursts, and Fast Radio Bursts. Yet their formation rate relative to ordinary neutron stars remains uncertain, often precluding direct comparisons with the rates of these extragalactic transients. Furthermore, magnetars have been recently shown to be evolutionarily related to other neutron star classes, complicating the estimate of the exact magnetar fraction within the neutron star population. We study the magnetar birth fraction in core-collapse supernovae using pulsar population synthesis of all isolated neutron star classes in our Galaxy, incorporating self-consistently the Galactic dynamical evolution, spin-down and magneto-thermal evolution. This approach allows us to derive strong constraints from small close-to-complete observational samples. In particular, looking at the age-limited young ($<$2\,kyr) neutron star population in the Milky Way we find 24 detected young neutron stars, with only 10 of them (41\%) being classical rotational powered pulsars, while the others (59\%) are either magnetars or central compact objects, the latter believed to be equally magnetically powered. We further compare the results with the nearby volume-limited class ($<$500\,pc) of X-ray Dim Isolated Neutron stars, old nearby magnetars. We conclude that the observed population of isolated neutron stars in the Galaxy can be reproduced only by assuming a core-collapse supernova rate larger than two, and a larger magnetar fraction than previously inferred. By assuming a bimodal initial magnetic field ($B_0$) distribution at birth, we find that the magnetar class peaks between $B_0\sim 1-2.5\times10^{14}$\,Gauss and represents on average $\sim50$\% of the entire neutron star population.
\end{abstract}

% MAIN TEXT (3500, 6 figures/tables, 50 refs).
%Extended data 10 more figure/tables.

%why it is important
Understanding the fraction of core-collapse supernovae (CC-SNe) that give birth to highly magnetic neutron stars, namely magnetars, rather than ordinary neutron stars is essential for constraining the physical pathways that shape the most extreme compact objects in the Universe. Magnetars have long been proposed as central engines capable of injecting large amounts of energy into their surrounding supernova ejecta, and are therefore prime candidates for powering superluminous supernovae (SL-SNe)\cite{Inserra2013}, as well as the plateau phases observed in many gamma-ray burst (GRB) afterglows \cite{Metzger2011, Lu_Zhang2014}. Moreover, the discovery of millisecond-duration fast radio bursts (FRBs) in the extragalactic sky \cite{FRB_review2022} has intensified interest in the formation rates of young magnetars in other galaxies, since their enormous magnetic fields and impulsive energy releases offer a compelling mechanism for generating such luminous radio transients. Establishing the birth rate of magnetars relative to normal pulsars is thus not only a key step in mapping the evolutionary outcomes of massive stars, but also in assessing whether the magnetar channel can statistically account for the observed population of related extragalactic high-energy and radio transients.

%What was known thus far
Previous efforts to quantify the fraction of CC-SNe that yield magnetars rather than ordinary pulsars have relied on a variety of methodologies. One recent line of work derives constraints from putting limits on the extragalactic rate of magnetar Giant Flares, large X-ray and gamma-ray flares powered by the magnetic energy and field instabilities of young magnetars (three detected thus far in our Galaxy). Giant Flares release a substantial fraction of the internal magnetic free energy of a magnetar in a single event. Hence their observed volumetric rate might in principle encode information about the magnetar formation fraction. However, the reliability of this approach remains limited by strong observational and theoretical biases: this method currently rests on a very small number of confidently identified extragalactic events, on the challenge of distinguishing them from cosmological short GRBs, and on the assumption of having controlled collimations fractions, luminosity functions and rates.

A second, independent strategy infers the magnetar fraction through population–synthesis modelling of the pulsar population. In this approach, neutron stars are evolved forward in time under assumptions for their initial spin period, magnetic–field distribution, and space velocities \cite{Beniamini2019, Sautron2025}. Model parameters are then tuned to reproduce the observed $P{-}\dot{P}$ diagram and the Galactic detection statistics. However, extending this framework to magnetars is intrinsically difficult: magnetars are not discovered through uniform, self-consistent surveys. In particular, young magnetars ($<2-5$\,kyr) are very bright and observable also at large distances, while medium and older systems are typically identified only when they enter an outburst phase. This results in a selection bias that is essentially unconstrained because the outburst rate is unknown and the physics behind inferring their rates are very model dependent. In addition, their persistent X-ray emission can be estimated only through magneto-thermal evolution models that incorporate detailed crustal microphysics, and such simulations have not yet been implemented in population–synthesis calculations, further limiting the reliability of inferred magnetar fractions. To date, estimates obtained with these methods suggest magnetar fractions in the range of 6$-$40\%, but they remain highly model-dependent and do not account for the magnetar evolutionary connection in the context of the full isolated neutron-star population.

%This value is consistent with the estimates of \citep{Beniamini2019}, who also used SNR ages as proxies for true ages, though their inferred range is broader than ours. Conversely, our inferred fraction is higher than that found by \citep{Sautron2025}, who obtained a magnetar birth fraction of 6–30\% using Bayesian inference only considering the observed magnetar population. Their approach, however, is more sensitive to observational biases, such as detection through transient outbursts.

% Magnetars are evolutionary connected with other NS classes, hence the problem is more general within the NS population.
On the other hand, estimating the magnetar fraction within the neutron star population cannot ignore their evolutionary links to other neutron-star subpopulations. Magnetars are thought to form a continuous evolutionary sequence with X-ray dim isolated neutron stars (XDINSs), central compact objects (CCOs), and high-$B$ radio pulsars, implying that the inferred birth rate of magnetars depends sensitively on how accurately population–synthesis models capture the transitions among these classes. In practice, this means that reliable rate estimates must incorporate these additional evolutionary pathways rather than treating magnetars in isolation.

% Mention in a couple of sentences what we do here?
In this work we use population synthesis simulations for all isolated Galactic neutron star classes taking into account their evolutionary relation using magneto-thermal evolutionary models for the first time. We constrain the magnetar birth fraction by comparing our simulations with neutron star sub-samples which are less biased observationally, and close to complete, focusing on the young ($<2$\,kyr) neutron stars, and the close-by ($<500$\,pc) population.

% All started with the observed population of your NSs
We start from considering all known Galactic neutron stars that have either a supernova remnant (SNR) age younger than 2\,kyr, or without a strong association, a characteristic age $\tau_c <2$\,kyr (which is known to be an upper limit on the true age; see Extended Data Figure\,\ref{fig:charact_vs_real_age} and Methods). We list them in Table\,\ref{tab:young_ns}, with the last column showing the source class, distinguishing between magnetars (which have exhibited magnetically powered bursts and outbursts), rotation-powered pulsars (RPPs; mostly powered by rotational energy dissipation), and CCOs (enigmatic thermally emitting neutron stars believed to be magnetars that experienced fallback accretion at birth, partially burying their magnetic fields). Neutron stars younger than 2\,kyr are expected to be extremely luminous across the radio, X-ray and gamma-ray bands regardless of class, which minimizes the likelihood that many such sources have been missed in the sample. 

%deriving CC rates in the Galaxy
From this observational census alone, two conclusions emerge without requiring further modelling. The number of observed young isolated neutron stars implies that the Galactic CC-SN rate must exceed one per century: the existence of 24 neutron stars younger than 2\,kyr is inconsistent with a lower rate, even if a small fraction of neutron stars were produced through alternative channels\cite{Hu_Zhang2025} and acknowledging existing observational biases. To independently check our assessment of the CC-SN rate in our Galaxy, we also studied a volume limited sample of core-collapse SNR within a 2\,kpc distance to derive an observational lower limit by independent means. We found 18 confirmed SNRs\cite{SamarSNRcatalog} within a 2\,kpc distance and $<$10\,kyr in age, 8 of which have a neutron star association (see Extended Data Table~\ref{tab:snr_2kpc}, Extended Data Figure~\ref{fig:age_cum} and Methods). We then extrapolated these numbers by simulating the observed population for the whole Galactic volume (taking into account proper motions and massive star distribution in the Galactic arms; see \cite{Ronchi2021} and Methods). This allow us to derive a lower limit on the CC-SN rate of $>2$ per century (see Methods for details).
Note that this is consistent with the most recent estimate of $1.63\pm0.46$ CC-SNe per century in the Milky Way \citep{Rozwadowska2021}.

%hence we are almost complete in our yung NS sample and more than half of them are magnetars
Consequently, assuming the plausible value of 1-3 CC-SNe per century the expected number of neutron stars younger than 2\,kyr in the Galaxy should be $<$60. Identifying 24 of them suggests that the young neutron star population is close to complete (missing $<$2/3), as anticipated from the strong multi-wavelength luminosity of such young objects. We also observe from the table that magnetars and CCOs together account for about 14 of the 24 young neutron stars, i.e., nearly half of the sample. 

% That said we need to do things properly with pop synthesis simulations 
However, to derive quantitative numbers and constraints on the magnetar fraction within the neutron star population in our Galaxy we need to perform detailed population synthesis simulations to take into account evolutionary scenarios between the different neutron star classes, and even more importantly the different observational biases involved.

To constrain the birthrate of magnetars while accounting for observational biases, we combine the observed young neutron star population with population–synthesis simulations using the {\tt ML-Poppyns} pulsar population synthesis code \citep{Ronchi2021, Graber2024, Pardo2025}. We perform 1000 simulations evolving neutron stars up to 2\,kyr, then evolving 100 of them until $3\times10^{7}$\,yr, including both dynamical and magneto–rotational evolution. Natal kicks are sampled from a log–normal distribution consistent with observed proper motions.

For the magneto-rotational evolution, we adopt initial distributions for both the birth spin period and the magnetic field. The former is drawn from a log-normal distribution centered at $\log_{10}(P/{\rm s}) = -0.67$ with a standard deviation of 0.55 \citep{Pardo2025} (see Methods). For the latter, we adopt a double log-normal distribution (see Extended Data \figurename~\ref{fig:init_mag}) to account for the existence of two distinct populations: magnetars and RPPs. The first log-normal distribution, representing standard radio pulsars, is centered at $\log_{10}(B/{\rm G}) = 13.09$ with a standard deviation of $0.5$ \citep{Pardo2025}. The second, representing magnetar-like sources, is centered at higher fields. The normalization of this second log-normal directly yields the magnetar birth fraction relative to the total neutron star population at birth. Therefore, our main focus is on inferring the value of the normalization that best matches the observations. We fix its standard deviation at 0.5 (see also Methods) and explore a range of values for the mean $B_0$ between: $7.5 \times 10^{13}$ Gauss and $10^{15}$ Gauss. The X-ray emission of the sample is computed self-consistently using magneto-thermal simulations, including thermal seed photons undergoing resonant cyclotron scattering and interstellar absorption (see Methods). To identify plausible values of the magnetar birth field, we compare the simulated and observed populations of both young magnetars (including CCOs in this class) and XDINSs. To quantify the magnetar fraction in our simulated population of young neutron stars and enable a direct comparison with the observed sample listed in Table \ref{tab:young_ns}, we classify sources as rotation-powered pulsar (RPP)–like if $B < 10^{13.5}$ G and $\dot{E}_\mathrm{rot} > 10^{36}$ erg s$^{-1}$ (consistent with the young pulsars we observe). Sources are classified as magnetar-like if $B > 10^{13.5}$ G and their observable X-ray flux exceeds $10^{-14}$ erg s$^{-1}$ cm$^{-2}$ (see Methods). For each assumed mean magnetic field, we compute the cumulative distributions of period, period derivative, and absorbed X-ray flux (see Extended Data Figure ~\ref{fig:cumulative_distributions}) and perform Kolmogorov–Smirnov (K–S) tests for each variable. The only initial magnetic fields that simultaneously satisfy these criteria lie in the $1\times10^{14}$~G and $2.5\times10^{14}$~G range. 

\begin{figure*}
    \centering
    \includegraphics[width=0.99\linewidth]{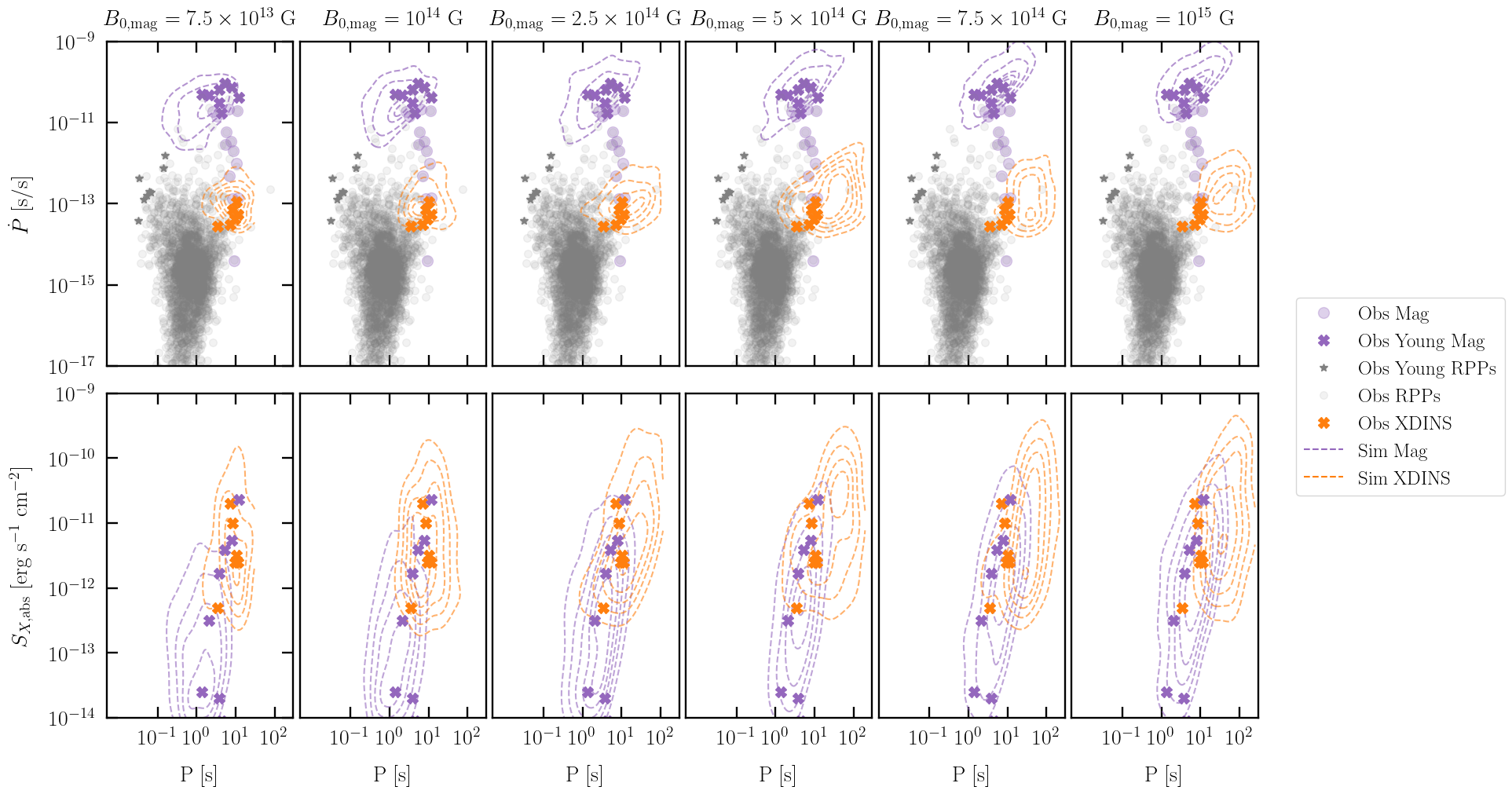}
    \caption{Simulated magnetars (purple contour lines) and XDINSs (orange contour lines) for different mean initial magnetar magnetic field strength at birth compared to the observed magnetars young population from Table~\ref{tab:young_ns} (purple crosses) and the observed XDINSs (orange crosses). The observed young RPPs from Table~\ref{tab:young_ns} are indicated with gray stars. The entire observed magnetar and \ac{RPP} populations are also shown as a reference with purple and gray circles respectively. The top row displays the $P-\dot{P}$ diagrams, while the bottom row shows the spin period versus absorbed X-ray flux ($S_{X,\rm abs}$). The contours indicate the density of detected magnetically powered neutron stars obtained from 100 simulations, filtered according to the criteria described in section "Inferring the magnetar birth fraction from population synthesis" of the Methods. }\label{fig:pop_sim_ppdot}
\end{figure*}

\begin{figure*}
    \centering
\includegraphics[width=0.6\textwidth]{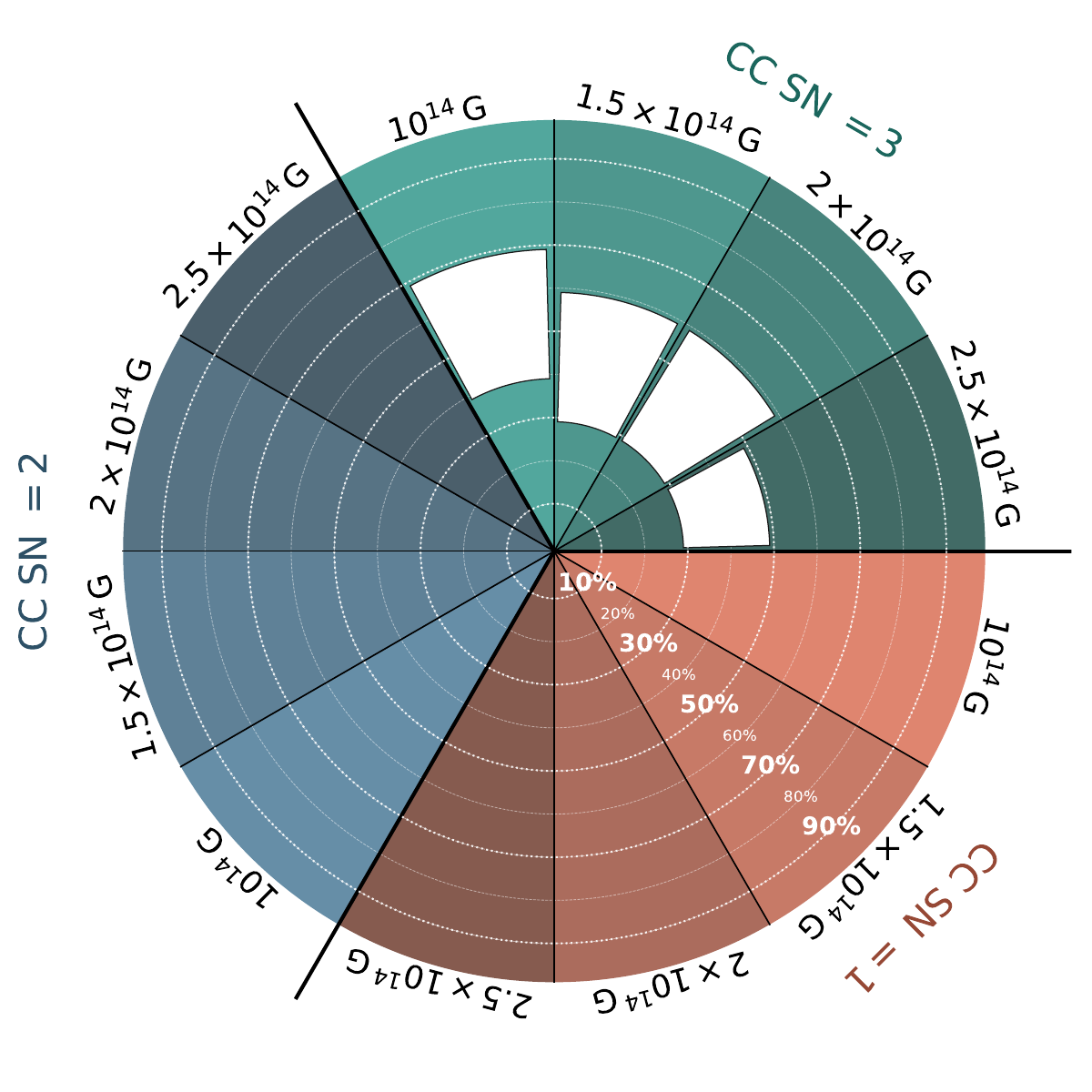}
\caption{Pie chart summarizing the inferred magnetar fractions for different \ac{CC-SN} rates and initial mean magnetar magnetic-field strengths. The chart is divided into three macro-sectors corresponding to a \ac{CC-SN} rate of 1 (brown), 2 (blue) and 3 (green), respectively. Each sector is then further divided into different magnetic field values representing the mean values of the magnetar initial magnetic field distribution. The parts of the sectors highlighted in white represent the inferred ranges of magnetar fractions at birth that are compatible with observations for each combination of \ac{CC-SN} rate and initial mean magnetar magnetic-field strength. For \ac{CC-SN} rate of 1 and 2 there is no combination of magnetic field distributions and magnetar fraction that are compatible with the observed population.
}
\label{fig:piechart}
\end{figure*}

A comparison of the simulated and observed populations (see \figurename~\ref{fig:pop_sim_ppdot}) shows excellent agreement in both $P$–$\dot{P}$ space and in the period–flux plane. We then perform 1000 simulations for each combination of CC-SN rate, magnetar birth fraction, and initial magnetar field strength (Extended Data \figurename~\ref{fig:pop_sin_sim}). For rates of one and two CC-SN per century, no combination of parameters simultaneously reproduces the observed numbers of both RPP-like and magnetar-like sources, as well as the close-by XDINSs, as expected. For three CC-SN per century, the observed population is matched with different magnetar fractions depending on the initial B-field distribution, ranging from $40-70$\% for a magnetar-like distribution peaking at $B\sim1\times10^{14}$\,G until $30-50$\% for fields peaking at $B\sim2.5\times10^{14}$\,G (see Figure\,\ref{fig:piechart} and Extended Data Figure\,\ref{fig:pop_sin_sim}).

Taken together, our findings support a CC-SN rate of roughly three isolated neutron stars per century and suggest that $\sim50$\% (averaging on the chosen initial magnetic field distributions) of them are born as magnetars. 
%This value is consistent with the estimates of \citep{Beniamini2019}, who also used SNR ages as proxies for true ages, though their inferred range is broader than ours. Conversely, our inferred fraction is higher than that found by \citep{Sautron2025}, who obtained a magnetar birth fraction of 6–30\% using Bayesian inference only considering the observed magnetar population. Their approach, however, is more sensitive to observational biases, such as detection through transient outbursts.
The high inferred birth fraction of magnetar-like neutron stars lends strong support to models in which these objects act as the central engines of a large percentage of extragalactic transients.

%\item To this we also add the XDINs limiting the PPS to 500pc and evolving until $\sim10^7$years, the oldest known XDINS.
%\item conclusions 

%%%%%%%%%%%%% FIGURE  %%%%%%%%%%%%% 
%%%%%%%%%%%%%%%%%%%%%%%%%%%%%%%%%%%%%%% 

%%%%%%%%%%%%% FIGURE  %%%%%%%%%%%%% 
%%%%%%%%%%%%%%%%%%%%%%%%%%%%%%%%%%%%%%% 

%-----------------------------TABLES--------------------
%------------------------------------------------------------------------------------------
\begin{table*}
\footnotesize
\setlength{\tabcolsep}{2.5pt}
\centering
\caption{Galactic neutron stars younger than 2 kyrs.}
\label{tab:young_ns}
\begin{tabular}{l l l l l l l l l l r}  
\hline\hline       
&SNR & NS name & SNR Age & dist & $\tau_{\rm c}$ &Period & $B_{\rm dip}$ & $\dot{E}_{\rm rot}$ &  Type\\ 
&  & & [kyr] & [kpc] & [kyr] & [s] & $10^{14}\,\mathrm{G}$ & $[\mathrm{erg}\,\mathrm{s}^{-1}]$ &  \\
\hline  
1& -- & Swift~J1818.0-1607 &  -- &  & 0.47 & 1.360 & 3.523 & $1.4\times 10^{36}$ & Mag (R,X) \\
2&G327.2-00.1 & 1E~1547-5408 & $0.8$\cite{Ranasinghe23} & 4–5 & 0.69 & 2.070 & 4.245 & $3.8\times 10^{35}$ & Mag (R,X) \\
3&G029.7-00.3 & PSR~J1846-0258 & $1.69$–$1.85$\cite{Leahy2018} & 5.3–5.9 & 0.73 & 0.330 & 0.654 & $1.4\times 10^{37}$ & Mag (X) \\
4&G042.8+00.6$^{**}$ & SGR~1900+14 & $110.0$\cite{Ranasinghe23} & $>2.8$ & 0.9 & 5.200 & 9.345 & $4.6\times 10^{34}$ & Mag (X) \\
5&G348.7+00.3 & CXOU~J1714-3810 & $0.65$–$16.8$\cite{Halpern2010} & 9.8–13.2 & 0.95 & 3.830 & 6.689 & $8 \times 10^{34}$ & Mag (X) \\
6&Crab & PSR~J0534+2200 & $0.97^{*}$ & 1.70–2.1 & 1.26 & 0.033 & 0.051 & $7.9 \times 10^{38}$ & RPP (R,X,$\gamma$) \\
7&G320.4-01.2 & PSR~B1509-58 & $1.9$\cite{Fang2010b} & 3.8–6.6 & 1.58 & 0.152 & 0.206 & $3\times 10^{37}$ & RPP (R,X,$\gamma$) \\
8&G292.2-00.5 & PSR~J1119-6127 & $4.2$–$7.1$\cite{Kumar2012} & 8.1–9.7 & 1.6 & 0.407 & 0.545 & $4.2\times 10^{36}$ & Mag (R,X) \\
9&  --  & SGR~1806-20 &  --  &  & 1.6 & 7.750 & 10.301 & $1.1\times 10^{34}$ & Mag (X)\\
10&G000.9+00.1 & PSR~J1747-2809 & $1.9$\cite{Ranasinghe23} & 8.34–10.0 & 1.9 & 0.052 & 0.038 & $7.7\times 10^{37}$ & RPP (R) \\
11&  --  & SGR~J1745-2900 & --  &  & 1.9 & 3.760 & 4.583 & $4\times 10^{34}$ & Mag (R,X) \\
12&G054.1+00.3 & PSR~J1930+1852 & $1.5$–$2.4$\cite{Leahy2018} & 4.1–7.2 & 2.89 & 0.137 & 0.137 & $2.1\times 10^{37}$ & RPP (R,X) \\
13&G027.4+00.0 & 1E~1841-045 & $0.75$–$2.1$\cite{Leahy2018}\cite{Borkowski2017} & 5.8–9.8 & 4.60 & 11.79 & 9.382 & $1.7 \times 10^{33}$ & Mag (X) \\
14&G021.5-00.9 & PSR~J1833-1034 & $1.55$–$1.8$\cite{Leahy2018} & 4.2–4.6 & 4.85 & 0.062 & 0.048 & $6\times 10^{37}$ & RPP (R,$\gamma$) \\
15&G000.9+00.1 & PSR~J1747-2809 & $1.9$\cite{Ranasinghe23} & 8.34–10.0 & 5.31 & 0.052 & 0.038 & $7.7\times 10^{37}$ & RPP (R) \\
16&G130.7+03.1 & PSR~J0205+6449 & $0.84^{*}$ &  & 5.37 & 0.066 & 0.048 & $4.8\times 10^{37}$ & RPP (R,X,$\gamma$)  \\
17&G012.8-00.0 & PSR~J1813-1749 & $1.2$\cite{Fang2010} & 4.0 & 5.58 & 0.045 & 0.032 & $9.9 \times 10^{37}$ & RPP (R,X)\\
18&G310.6-01.6 & PSR~J1400-6325 & $1.0$–$2.5$\cite{Reynolds19} & 5.0–10.0 & 12.7 & 0.031 & 0.015 & $9.2\times 10^{37}$ & RPP (R,X) \\
19&G011.2-00.3 & PSR~J1811-1925 & $1.4$–$2.4^{*}$\cite{Borkowski16} & 3.7–4.7 & 23.3 & 0.065 & 0.023 & $1.1\times 10^{37}$ & RPP (X) \\
20&G332.4-00.4 & 1E~161348-5055 & $2.0$–$4.4$\cite{Leahy2020} & 2.7–3.3 & -- & 24030 & -- & -- & Mag/CCO (X) \\
21&G111.7-02.1 & CXO~J2323+58 & $0.32$–$0.35$\cite{Suzuki2021} & 3.3–3.7 &  -- & --  &  -- &  -- & CCO (X)\\
22&G330.2+01.0 & CXO~J1601-513 & $<1$\cite{Tian2007} & $>$5.0 &  -- &  -- &  -- &  -- & CCO (X)\\
23&G347.3-00.5 & 1WGA~J1713-3949 & $1.8$–$2.4$\cite{Acero2017} & 0.9–1.1 &  -- &  -- &  -- &  -- & CCO (X) \\
24&G350.1-00.3 & XMMU~J1720-37 & $<0.7$\cite{Mayer21} & 4.5–9.0 &  -- &   --&   --&  -- & CCO (X) \\
\hline                 
\end{tabular}
(*) denote historical supernova age determinations. (**) a dubious association.
\end{table*}

\newpage

\renewcommand\thefigure{\arabic{figure}} 
\setcounter{figure}{0}
\setcounter{table}{0}
\renewcommand{\figurename}{Extended Data Figure}
\renewcommand{\tablename}{Extended Data Table}

\newpage
\newpage

%%%%%%%%%%%%%%%%%%%%%%%%%%%%%%%%%%%%%%%%%%%%%%%%%%%%%%%
% Article bibliography
\newcommand{\actaa}{Acta Astron.}   % Acta Astronomica
\newcommand{\araa}{Annu. Rev. Astron. Astrophys.}   % Annual Review of Astron and Astrophys
\newcommand{\areps}{Annu. Rev. Earth Planet. Sci.} % Annual Review of Earth and Planetary Science
\newcommand{\aar}{Astron. Astrophys. Rev.} % Astrononmy and Astrophysics Review
\newcommand{\ab}{Astrobiology}    % Astrobiology
\newcommand{\aj}{Astron. J.}   % Astronomical Journal
\renewcommand{\ac}{Astron. Comput.} % Astronomy and Computing
\newcommand{\apart}{Astropart. Phys.} % Astroparticle Physics
\newcommand{\apj}{Astrophys. J.}   % Astrophysical Journal
\newcommand{\apjl}{Astrophys. J. Lett.}   % Astrophysical Journal, Letters
\newcommand{\apjs}{Astrophys. J. Suppl. Ser.}   % Astrophysical Journal, Supplement
\newcommand{\ao}{Appl. Opt.}   % Applied Optics
\newcommand{\apss}{Astrophys. Space Sci.}   % Astrophysics and Space Science
\newcommand{\aap}{Astron. Astrophys.}   % Astronomy and Astrophysics
\newcommand{\aapr}{Astron. Astrophys. Rev.}   % Astronomy and Astrophysics Reviews
\newcommand{\aaps}{Astron. Astrophys. Suppl.}   % Astronomy and Astrophysics, Supplement
\newcommand{\baas}{Bull. Am. Astron. Soc.}   % Bulletin of the AAS
\newcommand{\caa}{Chin. Astron. Astrophys.}   % Chinese Astronomy and Astrophysics
\newcommand{\cjaa}{Chin. J. Astron. Astrophys.}   % Chinese Journal of Astronomy and Astrophysics (now RAA)
\newcommand{\cqg}{Class. Quantum Gravity}    % Classical and Quantum Gravity
\newcommand{\epsl}{Earth Planet. Sci. Lett.}    % Earth and Planetary Science Letters
\newcommand{\frass}{Front. Astron. Space Sci.}    % Frontiers in Astronomy and Space Sciences
\newcommand{\gal}{Galaxies}    % Galaxies
\newcommand{\gca}{Geochim. Cosmochim. Acta}   % Geochimica Cosmochimica Acta
\newcommand{\grl}{Geophys. Res. Lett.}   % Geophysics Research Letters
\newcommand{\icarus}{Icarus}   % Icarus
\newcommand{\ija}{Int. J. Astrobiol.} % International Journal of Astrobiology
\newcommand{\jatis}{J. Astron. Telesc. Instrum. Syst.}  % Journal of Astronomical Telescopes, Instruments, and Systems 
\newcommand{\jcap}{J. Cosmol. Astropart. Phys.}   % Journal of Cosmology and Astroparticle Physics
\newcommand{\jgr}{J. Geophys. Res.}   % Journal of Geophysics Research
\newcommand{\jgrp}{J. Geophys. Res.: Planets}    % Journal of Geophysics Research: Planets
\newcommand{\jqsrt}{J. Quant. Spectrosc. Radiat. Transf.} % Journal of Quantitiative Spectroscopy and Radiative Transfer
\newcommand{\lrca}{Living Rev. Comput. Astrophys.}    % Living Reviews in Computational Astrophysics
\newcommand{\lrr}{Living Rev. Relativ.}    % Living Reviews in Relativity
\newcommand{\lrsp}{Living Rev. Sol. Phys.}    % Living Reviews in Solar Physics
\newcommand{\memsai}{Mem. Soc. Astron. Italiana}   % Mem. Societa Astronomica Italiana
\newcommand{\maps}{Meteorit. Planet. Sci.} % Meteoritics and Planetary Science
\newcommand{\mnras}{Mon. Not. R. Astron. Soc.}   % Monthly Notices of the RAS
\newcommand{\nat}{Nature} % Nature
\newcommand{\nastro}{Nat. Astron.} % Nature Astronomy
\newcommand{\ncomms}{Nat. Commun.} % Nature Communications
\newcommand{\ngeo}{Nat. Geosci.} % Nature Geoscience
\newcommand{\nphys}{Nat. Phys.} % Nature Physics
\newcommand{\na}{New Astron.}   % New Astronomy
\newcommand{\nar}{New Astron. Rev.}   % New Astronomy Review
\newcommand{\physrep}{Phys. Rep.}   % Physics Reports
\newcommand{\pra}{Phys. Rev. A}   % Physical Review A: General Physics
\newcommand{\prb}{Phys. Rev. B}   % Physical Review B: Solid State
\newcommand{\prc}{Phys. Rev. C}   % Physical Review C
\newcommand{\prd}{Phys. Rev. D}   % Physical Review D
\newcommand{\pre}{Phys. Rev. E}   % Physical Review E
\newcommand{\prl}{Phys. Rev. Lett.}   % Physical Review Letters
\newcommand{\psj}{Planet. Sci. J.}   % Planetary Science Journal
\newcommand{\planss}{Planet. Space Sci.}   % Planetary Space Science
\newcommand{\pnas}{Proc. Natl Acad. Sci. USA}   % Proceedings of the US National Academy of Sciences
\newcommand{\rpp}{Rep. Prog. Phys.}
\newcommand{\procspie}{Proc. SPIE}   % Proceedings of the SPIE
\newcommand{\pasa}{Publ. Astron. Soc. Aust.}   % Publications of the Astron. Soc. of Australia
\newcommand{\pasj}{Publ. Astron. Soc. Jpn}   % Publications of the Astron. Soc. of Japan (note no full stop following Jpn)
\newcommand{\pasp}{Publ. Astron. Soc. Pac.}   % Publications of the Astron. Soc. of the Pacific
\newcommand{\raa}{Res. Astron. Astrophys.} % Research in Astronomy and Astrophysics (formerly CJAA)
\newcommand{\rmxaa}{Rev. Mexicana Astron. Astrofis.}   % Revista Mexicana de Astronomia y Astrofisica
\newcommand{\sci}{Science} % Science
\newcommand{\sciadv}{Sci. Adv.} % Science Advances
\newcommand{\solphys}{Sol. Phys.}   % Solar Physics
\newcommand{\sovast}{Soviet Astron.}   % Soviet Astronomy
\newcommand{\ssr}{Space Sci. Rev.}   % Space Science Reviews
\newcommand{\uni}{Universe} % Universe
\bibliography{biblio.bib}
% ADD Here in bibtex format from the compile bbl file only those relative to the Article, and after the Methods section those relative to the Methods.
%%%%%%%%%%%%%%%%%%%%%%%%%%%%%%%%%%%%%%%%%%%%%%%%%%%%%%%
% METHODS

\newpage
\begin{methods}

\subsection{Observational data}\label{sec:sample}
\label{sec:obs_data}
In Table~\ref{tab:young_ns}, we list neutron stars whose ages are consistent with being younger than 2 kyr, along with any associated supernova remnants (SNRs). The age of each neutron star is determined either from its association with a \ac{SNR}, when such an association exists, or from its characteristic age, which serves as a proxy for the true age in the absence of a known remnant. For supernova-based ages, estimates are typically given with an associated uncertainty. Therefore, we include sources whose estimated ages are compatible with the 2 kyr threshold within their error margins. For each source, we provide the estimated age and distance of the supernova, as well as the neutron star’s characteristic age, period, and dipolar magnetic field strength. The data for magnetars and \ac{CCO} were taken from \cite{Rea2025} and \cite{Mayer21}, respectively. For the \ac{RPP}, we used data from the ATNF Pulsar Catalogue \citep{Manchester2005}\footnote{\url{https://www.atnf.csiro.au/research/pulsar/psrcat/}, v2.6.0}. References for the \ac{SNR} ages and distances were initially taken from \cite{Giles12}\footnote{\url{http://snrcat.physics.umanitoba.ca}}. 
However, each age was individually reviewed and updated to reflect the most recent estimates available in the literature. As a result, some supernova ages may differ from those reported in \cite{Giles12}. The specific studies used for each updated age are listed in the table. In the table, we also include the classification of the associated neutron star, grouped into three categories: Magnetars, \ac{CCO}s and \ac{RPP}s. 
For pulsars and magnetars, we additionally indicate the wavelength band in which pulsations were detected, i.e., radio (R), X-ray (X), and/or gamma-ray ($\gamma$). For G042.8+00.6, although the SNR age exceeds 2 kyr and the association with SGR 1900+14 is uncertain, we adopt an age younger than 2 kyr for the magnetar. This assumption is motivated by its characteristic age of 0.9 kyr, which is expected to overestimate the true age (see the dedicated section on the characteristic ages below). A similar reasoning is applied to PSR J1119-6127, associated with G292.2-00.5. Finally, we excluded neutron star–SNR associations with unreliable age estimates, such as G338.3-00.0 and G333.9+00.0 from the dataset.

\subsection{Magneto-rotational evolution}

To model the spin evolution of neutron stars, we adopt the magnetospheric torque prescription described in \cite{Graber2024}. In this framework, the spin period $P$ and inclination angle $\chi$ between the rotation axis and the magnetic axis, evolve under the influence of plasma-filled magnetospheres, and their time evolution is obtained by solving the coupled differential equations \citep{Spitkovsky2006, Philippov2014}:
\begin{align}
\dot{P} &= \frac{\pi^2}{c^3} \frac{B^2 R^6}{I P} \left( \kappa_0 + \kappa_1 \sin^2 \chi \right), \\
\dot{\chi} &= -\frac{\pi^2}{c^3} \frac{B^2 R^6}{I P^2} \left( \kappa_2 \sin\chi \cos\chi \right),
\end{align}
where $c$ is the speed of light, $R \approx \unit[11]{km}$ is the neutron-star radius, and $I \simeq 2 M R^2 / 5 \approx \unit[1.36 \times 10^{45}]{g \, cm^{2}}$ is the stellar moment of inertia (for a fiducial mass $M \approx \unit[1.4]{M_{\odot}}$). For realistic pulsars surrounded by plasma-filled magnetospheres, we choose $\kappa_0 \simeq \kappa_1 \simeq \kappa_2 \simeq 1$.

We also incorporated magnetic field decay to accurately capture the long-term magneto-rotational evolution of neutron stars. The evolution of the dipolar magnetic field strength in neutron stars is influenced by both the Hall effect and ohmic dissipation in the crust \citep{Vigano2013}. This decay is particularly significant for strongly magnetized neutron stars with fields above $~10^{13}$ G, making it relevant for a considerable fraction of our simulated neutron star population. We compute the dipolar magnetic field at time $t$ by parameterizing the results from 2D magneto-thermal numerical simulations \citep{Vigano2021, Graber2024}.\\
\subsection{Characteristic age bias}

For our analysis, we need to evaluate whether our sample of young neutron stars is complete or whether a population of sources may be missing due to selection effects. One possible bias arises from excluding neutron stars with characteristic ages greater than 2 kyr and no association with a \ac{SNR}, thereby lacking more reliable age estimates. 
The characteristic age,  $\tau_{\rm c}$ , is defined as:
\begin{equation}
    \tau_{\rm c} =  \frac{P}{2\dot{P}},
\end{equation}
where $P$ is the neutron star spin period and $\dot{P}$ its first time derivative, representing the spin-down rate. The derivation of the $\tau_c$ assumes a braking index of three for pure magnetospheric dipole braking, a constant magnetic field, and an initial spin period much shorter than the current period. These assumptions do not always hold, and deviations can lead to substantial overestimation of a pulsar’s true age \citep{Gourgouliatos2018}. As a result, some young neutron stars may be underrepresented in our sample.

To quantify this effect, we rely on the magneto-rotational simulations presented in the previous section, which self-consistently yield both the true and characteristic ages of neutron stars. In \figurename~\ref{fig:charact_vs_real_age}, we compare the characteristic age and the true age of neutron stars, for different initial magnetic fields and spin periods. We consider two configurations: one in which the magnetic field decays over time, and another in which it remains constant. In the constant-field case (left panels of \figurename~\ref{fig:charact_vs_real_age}) at the beginning of the evolution, $\tau_c$ tends to overestimate the true age for neutron stars with typical pulsar-like magnetic fields. This discrepancy arises because the spin period is still close to its initial value, making the characteristic age unreliable. However, for high magnetic fields, such as those typical for magnetars, the characteristic and true ages are nearly identical at early times. This is because the strong magnetic field leads to a rapid spin-down due to a large $\dot{P}$, making $\tau_c$ a good proxy for the true age in young magnetars. When magnetic field decay is included (see right panels of \figurename~\ref{fig:charact_vs_real_age}), the evolution of the ratio $\tau_c / t$ initially resembles the constant-field case. However, as the field decay becomes significant, typically around $t \sim 10^5$ years, $\tau_c$ once again becomes an upper limit on the true age. This effect is more pronounced in magnetars, where the magnetic field decays faster than in RPPs \citep{Dehman2023, Vigano2013,Vigano2021}. 

Additionally, we explore a scenario in which $\dot{P}$ undergoes a sudden increase by a factor of 10 at $t = 1$ kyr. This mimics the variability in $\dot{P}$ observed in magnetars, which can be caused by magnetospheric changes or bursts of activity \citep{Dib2012}. This evolution is not shown in \figurename~\ref{fig:charact_vs_real_age}, as it introduces only a transient jump in the characteristic age $\tau_c$, which briefly underestimates the true age. However, $\tau_c$ quickly realigns with the true age because, at this stage of evolution, the spin period is already much larger than its initial value. As a result, this sudden change does not significantly affect the behavior depicted in \figurename~\ref{fig:charact_vs_real_age}.

The fact that the characteristic age overestimates the true age, is also reflected in \figurename~\ref{fig:characte_vs_real_b_dip}. The figure shows the characteristic age versus the age estimated from the associated \ac{SNR}s, including the corresponding uncertainties, for neutron stars whose SNR ages are lower than 10 kyr within errors \cite{Giles12}. The color scale represents the dipolar magnetic field strength. For most sources, the characteristic age exceeds the SNR age (i.e. points lie below the diagonal dashed line). However, for high-magnetic-field neutron stars, the two age estimates are more closely aligned, with points lying nearer to the diagonal. This behavior is consistent with \figurename~\ref{fig:charact_vs_real_age}, where the characteristic age is shown to be a reliable proxy for the true age in young strongly magnetized sources. We note that a few exceptions exhibit the opposite trend, with characteristic ages lower than the SNR ages (sources above the dashed diagonal line). These cases correspond to high magnetic-field neutron stars: PSR J1846-0258, PSR B1509-58, and SGR 1627-41. This discrepancy may be caused by uncertainties in the distance to the supernova, and hence in the inferred age. It may also arise if the true braking index is less than 3, and if the spin-down is not solely due to dipolar magnetospheric braking. For example, for the magnetar PSR~J1846$-$0258, there are conflicting age estimates in the literature: \cite{Leahy2008} argued for a smaller distance than in earlier works, implying a Sedov age consistent with the spin-down age of the neutron star. In the case of B1509$-$58 (associated with SNR~G320.4$-$01.2), the SNR and spin-down ages are more consistent, though the SNR age remains slightly larger. This small discrepancy could be accounted for if the braking index is less than three \citep{Camilo2002}. The characteristic age estimate of $2$ kyr for SGR 1627–41 should be regarded with caution, as it is based on a period derivative inferred from only two observational epochs, and magnetar spin-down rates are known to exhibit significant variability \cite{Esposito2009}.

\subsection{Core-collapse rate inferred from observations}
To robustly estimate the true fraction of magnetars in the Galaxy, a reliable determination of the \ac{CC-SN} rate is required. While constraining this rate is not the focus of this work, we assess whether the observed rate of SNRs with associated neutron stars is consistent with the commonly adopted estimate of 1–2 supernovae per century \citep{Rozwadowska2021}. To this end, we adopt two complementary strategies, both of which yield lower limits due to sample incompleteness.

First, we derive a hard lower limit by directly counting young SNRs with confirmed neutron star associations. In the second approach, we scale the observed local birthrate to the entire Galaxy by assuming a spatial distribution model for the \ac{SNR}s.

To derive a conservative lower limit from observations, we select from the catalog all \ac{SNR}s with neutron star associations and estimated ages younger than 10 kyr \cite{Giles12}. We focus specifically on SNRs with confirmed neutron star associations to ensure that the remnants in our sample originate from \ac{CC-SN}e.

In \figurename~\ref{fig:age_cum}, we present the cumulative age distribution of \ac{SNR}s in our selected sample. The green shaded region represents the distribution derived from the upper and lower bounds of the age estimates, capturing the uncertainties in individual SNR age determinations. For reference, we overplot a line corresponding to a constant supernova birthrate of 0.6 events per century. This indicates that, based on observational data alone, a minimum birthrate of 0.6 supernovae per century is required.

It is important to emphasize that this sample is not complete, owing to well-known selection effects in the detection of \ac{SNR}s\citep{Green2009}. Older SNRs, in particular, may have expanded and dissipated into the interstellar medium, reducing their surface brightness below current detection thresholds. This effect is reflected in the decline in the number of observed SNRs with increasing age above 3 kyrs, as shown in \figurename~\ref{fig:age_cum}. Given that the star formation rate has been approximately constant over the past 10 Myr \citep{Soler2023}, this decline highlights the incompleteness of the sample, likely driven by observational biases. Conversely, very young SNRs can remain unidentified due to their small angular sizes \citep{Green2009}. Nonetheless, the green distribution in \figurename~\ref{fig:age_cum} provides a conservative lower limit on the Galactic \ac{CC-SN} rate and, by extension, a lower bound on the neutron star birthrate.\\
The detectability of \ac{SNR}s is moreover strongly dependent on their distance, with more distant remnants being significantly harder to identify than nearby ones. This observational bias further affects the completeness of our sample. The impact of this distance-dependent selection effect is illustrated in \figurename~\ref{fig:spatial_dist_snr}, which shows the spatial distribution of SNRs in our sample in Galactocentric Cartesian coordinates, projected in the top-down (XY) and edge-on (XZ) views of the Galaxy, respectively. For comparison, we also overlay a simulated model of the Galaxy, including the spiral arm structure, built using the dynamical framework described below. We distinguish between \ac{SNR}s with associated neutron stars (stars) and those without (circles), in order to explore potential additional observational biases related to neutron star detectability. We also include error bars representing the uncertainties on the estimated distances for each SNR in the sample. As expected, the distribution of \ac{SNR}s approximately traces the spiral arms of the Galaxy, reflecting the locations of their progenitors. Massive OB stars are concentrated in star-forming regions. There is a clear absence of observed SNRs on the far side of the Galaxy, highlighting the strong observational bias against detecting distant remnants. In the edge-on view, the remnants are tightly confined to the Galactic plane, consistent with expectations: the short lifespans of OB stars prevents them from migrating far from their birth sites in the disk before exploding in \ac{CC-SN}e.
\subsection{Dynamical evolution}
We employ a unified dynamical evolution framework for both components of our analysis: the volumetric inference of the \ac{CC-SN} rate and the population-synthesis calculations used to constrain the magnetar fraction. Although the method was originally designed for neutron stars, it can be adapted to \ac{SNR}s by simply disabling the natal kick velocities. For a full description of the method, we refer the reader to \cite{Ronchi2021}. Initial positions are sampled in Galactocentric cylindrical coordinates $(r, \phi, z)$, adopting the Sun’s location as $r = \unit[8.3]{kpc}$, $\phi = \pi/2$, and $z = \unit[0.02]{kpc}$ \citep{Pichardo2012}.

The spatial distribution of OB stars, the progenitors of core-collapse supernovae, provides the underlying model from which we sample the initial positions of \ac{SNR}s, and thus the birth locations of neutron stars. In particular, we assume that the $z$-distribution follows an exponential disk profile \citep{Wainscoat1992}, drawing samples from a probability density function of the form:
 %----------------------------------------------------------------------------------------------
\begin{equation}
	\mathcal{P}(z) = \frac{1}{h_{\rm c}} \exp\left(-\frac{ \lvert z \rvert}{ h_{\rm c} } \right).
\end{equation}

We set the characteristic scale height to $h_{\rm c} = \unit[0.18]{kpc}$ \citep{Ronchi2021}, consistent with the vertical distribution of young massive stars in the Milky Way \citep{Li2019}. To ensure a symmetric distribution relative to the Galactic plane, we assign each star a $z$-coordinate drawn from this scale height, with its sign randomised to populate both sides of the plane evenly.

To determine the location in the Galactic disk, we assume a radial distribution model based on the distribution of observed \ac{SNR}s in the Galaxy \citep{Verberne2021}. Specifically, we use their Equation 9:
\begin{equation}
    f(r) = \exp \left( -\beta \frac{r - r_{\odot}}{r_{\odot}} \right).
\end{equation}
where $r_{\odot}= 8.3$ kpc is the distance of the Sun in galactocentric coordinates. For the model of the spiral arms, we assume a logarithmic shape function
which gives the azimuthal coordinate $\phi$ as a function of the distance from the Galactic center:
\begin{equation}
\label{eq:spiral_arms}
\phi(r) = k \ln\left( \frac{r}{r_0} \right) + \phi_0.
\end{equation}
Our model parameters, i.e., the winding constant $k$, the inner radius $ r_0$, and the inner angle $\phi_0$ are reported in Table 1 of \cite{Ronchi2021} and evaluated
from Table 1 in \cite{Yao2017} in order to match
the same functional form as defined in Eq. \ref{eq:spiral_arms}.

Assuming the Galaxy’s spiral arms rotate as a rigid structure with a period of $250$ Myr, we can estimate the birth angular position of each neutron star based on its age and assuming that the Galaxy rotate clockwise \citep{Vallee2017, Skowron2019}.\\

We evolve the positions and velocities in time by solving the Newtonian equation of motion \citep{Ronchi2021}. For this purpose, we assume that each star initially follows the azimuthal orbital velocity of its progenitor, which depends on the Galactic gravitational potential, $\Phi_{\rm MW}$:
\begin{equation}
	\boldsymbol{v}_{\rm orb} = \sqrt{ r \, \frac{\partial \Phi_{\rm MW} \left( r,z \right)}{\partial r} } \, \hat{\boldsymbol{\phi}},
\end{equation}

The gravitational potential, $\Phi_{\rm MW}$, is expressed as the sum of four components: the nucleus, $\Phi_{\rm n}$, the bulge, $\Phi_{\rm b}$, the disk, $\Phi_{\rm d}$, and the halo, $\Phi_{\rm h}$ \citep{Ronchi2021}.
When this dynamical framework is applied to neutron stars (rather than \ac{SNR}s), we additionally include natal kick velocities. These kicks, produced by asymmetries in the supernova explosion \citep{Coleman2022, Janka2022}, are drawn from a log-normal distribution with mean $\mu = 5.6$ and standard deviation $\sigma = 0.68$, consistent with the observed proper motions of isolated neutron stars \citep{Disberg2025}.

\subsection{Volumetric scaling of the core-collapse supernova rate}
To refine the estimate of the Galactic CCSN rate, we perform a volumetric scaling of the observed SNR population. Specifically, we focus on remnants located within a radius of 2 kpc from the Sun and extrapolate this observed number to the entire Galaxy, assuming a model for the spatial distribution. Although the 2 kpc radius is somewhat arbitrary, it is large enough to include a sufficient number of SNRs for analysis, but not so large that distance uncertainties dominate the sample. These remnants are listed in \tablename~\ref{tab:snr_2kpc} and correspond to those enclosed within the circle in \figurename~\ref{fig:spatial_dist_snr}.

We distinguish between sources with reliable distance and age estimates and those with greater uncertainties or dubious associations in \tablename~\ref{tab:snr_2kpc}. To estimate the Galactic SNR population, we conservatively assume completeness within this 2 kpc radius and restrict our analysis to the five SNRs with the most robust properties (i.e., the first five entries in \tablename~\ref{tab:snr_2kpc}). Although \mbox{SNR G263.9$-$03.3} (Vela pulsar) has a large uncertainty in its age, we include it in the sample since its associated neutron star has a characteristic age of 11 kyr. Given that $\tau_c$ overestimate the true age, particularly for low magnetic field pulsars, we consider it reasonable to assume this remnant is younger than 10 kyr.

To infer the total number of remnants in the Galaxy, we apply the dynamical framework described above to evolve a synthetic population of SNRs up to 10 kyr, matching the age range of the observed sample. We then estimate the expected number of supernovae in the Galaxy by simulating batches of 100 explosions, evolve their positions for 10 kyr, and count how many fall within 2 kpc of the Sun. When the number of simulated SNRs within this volume approaches five, we reduce the batch size to one to avoid overshooting. The simulation stops once five SNRs are found within the target radius. We repeat this process $500,000$ times to account for stochastic variation and low-number statistics, enabling a robust estimate of the Galactic SNR birthrate and its uncertainty. We assume the sample of SNRs younger than 10 kyr within 2 kpc is complete. However, this assumption may not hold due to the observational biases mentioned above. Thus, our birthrate estimate should be regarded as a lower limit. The resulting birthrate is shown in orange in \figurename~\ref{fig:age_cum}, with the light and dark orange bands indicating the 1$\sigma$ and 3$\sigma$ intervals, respectively, and the dashed orange line marking the median. Our estimated Galactic \ac{CC-SN} rate is $2.01^{+1.90}_{-0.96}$ events per century, consistent with previous estimates \cite{Rozwadowska2021}.

To test the robustness of our result, we repeated the analysis using different assumptions for the Sun’s position and the scale height $h_c$, which may vary with Galactocentric distance \citep{Li2019}. We adopted two Sun–Galactic center distances (8.0 and 8.3 kpc) and two scale heights ($h_c = 132$ pc and $h_c = 180$ pc). The resulting birthrate estimates were consistent across all configurations, indicating that our conclusions are not sensitive to these parameter choices.

\subsection{Population synthesis}

To constrain the magnetar birthrate while accounting for observational biases, we combine the observed young neutron star population with population synthesis simulations. Specifically, we perform isolated neutron star population synthesis in which we vary two free parameters: the fraction of magnetars at birth and the \ac{CC-SN} rate, which sets the total number of stars in the simulations. The allowed parameter space is defined by configurations that reproduce at least the observed numbers of both magnetically powered and rotation-powered neutron stars within the Galactic young neutron star population, namely nine magnetars and eight radio rotation-powered neutron stars (see Table~\ref{tab:young_ns}).\\ We simulated neutron star ages drawn from a uniform distribution with a maximum age of 2~kyr, consistent with the age range of the observed young neutron star population listed in Table~\ref{tab:young_ns}. We performed both the dynamical and magneto-rotational simulations described in the previous sections. Since the simulations were evolved for 2 kyr only, the neutron stars had insufficient time to move significantly within the Galaxy, and therefore the results  presented here are independent of the assumed kick velocity distribution.\\
A robust estimate of the magnetar birth fraction should be consistent not only with the youngest neutron stars but also with all other isolated neutron star classes within the population. For this reason, we start extending our analysis to the \ac{XDINS}. These sources have been proposed to represent evolved magnetars that were born with initial magnetic fields of the order of $10^{14}$~G, which have since decayed and can be explained within the framework of magneto-thermal evolution models \citep{Vigano2013}.

For this analysis, we use the observational sample of XDINSs compiled by \cite{Potekhin2020} and perform population synthesis simulations in which again we vary the magnetar fraction, the \ac{CC-SN} rate, and the initial magnetic field distribution of magnetars, following the set-up described. The simulations are evolved up to a maximum age of $3\times10^{7}$~yr, consistent with the characteristic age of the oldest known XDINS ($\approx 1.7\times10^{7}$~yr \cite{DeGrandis2022}).  \\

For the magneto-rotational evolution, we adopt initial distributions for both the birth spin period and magnetic field. The initial spin period is drawn from a log-normal distribution centered at $\log_{10}(P/{\rm s}) = -0.67$ with a standard deviation of $0.55$ \citep{Pardo2025}. We assume that both magnetars and rotation-powered pulsars share this same birth period distribution. This choice is motivated by the fact that magnetar spin periods evolve extremely rapidly due to their strong fields; after only $\sim 2$ kyr, the spin period has already lost memory of its initial value, effectively erasing any potential differences in spin period at birth.\\
For the magnetic field at birth, we adopt a double log-normal distribution (see \figurename~\ref{fig:init_mag}) to account for the existence of two distinct populations: magnetars and rotation-powered pulsars.  This assumption relies on the fact that a single log-normal distribution cannot reproduce the observations of both rotation-powered pulsars and magnetars, as demonstrated by earlier works \citep{Gullon2015, Sautron2025}, and can be physically motivated by the hypothesis that these neutron-star classes arise from progenitors with different properties or that experience different magnetic-field amplification mechanisms during core collapse.
The first log-normal distribution, representing standard radio pulsars, is centered at $\log_{10}(B/{\rm G}) = 13.09$ with a standard deviation of $0.5$ \citep{Pardo2025}. The second, representing magnetars, is centered at higher fields. The normalization of this second log-normal directly yields the magnetar birth fraction relative to the total neutron star population at birth. Therefore, our main focus is on inferring the value of the normalization that best matches the observations. We fix its standard deviation at 0.5 and explore a range of values for the mean between $7.5 \times 10^{13}$ G and $10^{15}$ G. To identify plausible values of the magnetar birth field, we compare the simulated and observed populations of both young magnetars and XDINSs. For each assumed mean magnetic field, we compute the cumulative distributions of period, period derivative, and absorbed flux (see \figurename~\ref{fig:cumulative_distributions}) and perform Kolmogorov–Smirnov (K–S) tests for each variable. The resulting K–S statistics and $p$-values are reported in Extended Data Tables~\ref{tab:ks_magnetars} and \ref{tab:ks_xdins}. Values highlighted in grey correspond to cases where the K–S $p$-values do not allow us to reject the null hypothesis at the 99\% level, indicating that the simulated and observed distributions are statistically consistent.
For both magnetars and XDINSs, the only initial magnetic fields that simultaneously satisfy these criteria lie in the range $1.5\times10^{14}$–$2\times10^{14}$~G. However, since constraining the magnetar birth-field distribution is not the primary focus of this work (and it is studied in detail in Ronchi, Pardo, Graber \& Rea 2026 in prep that will be submitted soon), and in order to maintain a conservative estimate of the magnetar birth fraction, we also include the neighbouring values $1\times10^{14}$~G and $2.5\times10^{14}$~G in our analysis. Although the former fails to reproduce the observed magnetar population and the latter fails for the XDINS sample, including these endpoints ensures that our inferred magnetar fraction brackets the full plausible range allowed by our discrete sampling of the parameter space.

\subsection{X-ray emission}
Having evolved the neutron stars in our sample dynamically and magneto rotationally, we model their X-ray thermal emission using the same magneto-thermal simulations employed for the magnetic field evolution. These simulations solve the coupled evolution of the magnetic field and temperature \citep{Vigano2021}. The resulting surface temperature profile is then used to compute the total luminosity at a given age under the assumption of blackbody emission. Our models also account for resonant cyclotron scattering, which produces the characteristic non-thermal tail observed in magnetar spectra \citep{Lyutikov2006,Rea2008}. This component arises as the thermal seed photons emitted from the surface are up-scattered via resonant Compton interactions with electrons gyrating along the magnetic field lines in the magnetosphere.
To compute the spectrum distorted by resonant cyclotron scattering we used the simplified semi-analytical 1D model described in \cite{Lyutikov2006}.
In this model it is assumed that the seed thermal photons propagate only in the radial direction and interact with the magnetospheric electron-positron plasma where particles are assumed to have a top-hat thermal velocity distribution centered at zero and extending up to velocities $\pm \beta_{\rm T}$. 
The effectiveness of the process is quantified by the resonant scattering optical depth $\tau_{\rm res}$ which can be estimated as \citep{Lyutikov2006, Rea2008}:
%-------------------------------------
\begin{align} \label{eq:tau_res}
    \tau_{\rm res} \sim \tau_0 (1 + \cos^2{\theta}),
\end{align}
%-------------------------------------
where $\theta$ is the angle between the magnetic field lines and the direction of propagation of the photons and:
%-------------------------------------
\begin{align} 
    \tau_0 = \frac{\pi^2 e^2 n_e r}{3 m_e c \omega_B}.
\end{align}
%-------------------------------------
For simplicity it is assumed that photons propagate parallel to the magnetic field lines, either away or towards the star, hence in Eq.~\eqref{eq:tau_res}, $\cos \theta = \pm 1$ and $\tau_{\rm res} = 2 \tau_0$ \citep{Lyutikov2006, Rea2008}.
Therefore this model depends on two free parameters, i.e., $\tau_0$ and $\beta_{\rm T}$. We assumed that these two parameters depends on the dipolar magnetic field strength $B$ of the neutron star according to the following scaling relations \citep{Gullon2015}:
%-------------------------------------
\begin{align} 
    \tau_0 &= \begin{cases} 0.001  \quad {\rm if} \, \unit[B \leq 10^{13}]{G} \\ \frac{B}{\unit[10^{14}]{G}} \quad {\rm if} \, \unit[B > 10^{13}]{G} \end{cases}
\end{align}
%-------------------------------------
%-------------------------------------
\begin{align} 
    \beta_{\rm T} = \begin{cases} 0.001  \quad {\rm if} \, \unit[B \leq 10^{13}]{G} \\ 0.3 \quad {\rm if} \, \unit[B > 10^{13}]{G}. \end{cases}
\end{align}
%-------------------------------------
These approximated relations roughly reproduce the correlations between $\tau_0$, $\beta_{\rm T}$ and $B$ \citep{Rea2008}.
With this in hand we can compute the scattering probabilities and the resulting intensity of the resonantly scattered spectrum $I_{\rm RCS}(E)$, where $E$ denotes the photons energy \citep{Lyutikov2006}. By knowing the distance $d$ from the neutron star we can derive the intrinsic flux density as:
%-------------------------------------
\begin{align} \label{eq:intrinsic_flux}
    S_{\rm X}(E) = \pi \left( \frac{R_{\infty}}{d} \right)^2 I_{\rm RCS}(E),
\end{align}
%-------------------------------------
where $R_{\infty}$ is the neutron star radius as seen by a distant observer after applying general relativistic corrections.

Interstellar absorption due to photoionisation is then applied to estimate the flux reaching Earth \citep{Wilms2000}. 
The absorbed flux is given by:
%-------------------------------------
\begin{align} \label{eq:absorbed_flux}
    S_{\rm X, obs}(E) = e^{-\sigma_{\rm ISM}(E) N_{\rm H}} S_{\rm X}(E),
\end{align}
%-------------------------------------
where $\sigma_{\rm ISM}(E)$ represents the energy-dependent effective absorption cross section of the \ac{ISM} and $N_{\rm H}$ denotes the hydrogen column density along the line of sight of a given neutron star.
To compute $\sigma_{\rm ISM}(E)$ we only take into account the neutral atomic gas and use the routines from \citep{Balucinska-Church1992}.
For a given neutron star with known equatorial sky coordinates and distance $d$ the value of $N_{\rm H}$ can be estimated using the reddening map of the Galaxy and the calibration factor provided by \citep{Doroshenko2024}.
For a more detailed description of the procedure used to compute X-ray fluxes, we refer the reader to Ronchi, Pardo-Araujo, Graber \& Rea (2026 in preparation).

\subsection{Inferring the magnetar birth fraction from population synthesis}
Once the fully evolved population is obtained, we distinguish two classes of neutron stars among the young population ($<2$ kyrs): rotation-powered and magnetically powered. A neutron star is classified as rotation-powered if its dipolar magnetic field is lower than $10^{13.5}$ G, and it is detected if its rotational energy, $|\dot{E}_{\rm rot}| = 4 \pi^2 I \dot{P} / P^3$, exceeds $10^{36}$ erg s$^{-1}$. This criterion is motivated by the observed young RPP population listed in Table~\ref{tab:young_ns}. We further assume a beaming fraction of 0.5, so that only $50\%$ of the population satisfying the above criteria is detectable given that beams will intercept our line of sight. This relatively large beaming fraction compared to that typically inferred in pulsar population studies is justified by considering an inverse relation between beaming aperture and period \citep{Gangadhara2001}. In particular, for young radio pulsars, the period has not yet evolved significantly, resulting in a larger beaming aperture than in older populations, where it is typically $\sim 10\%$ \citep{Gullon2015, Graber2024}. We classify a neutron star as magnetically powered if its magnetic field exceeds $10^{13.5}$ G, and it is considered detected if its X-ray flux exceeds $10^{-14}$ erg s$^{-1}$ cm$^{-2}$. Similar to the criteria applied for rotation-powered neutron stars, these thresholds are motivated by the observed properties of the stars listed in \tablename~\ref{tab:young_ns}. For the old neutron star population, we identify as XDINS all those sources with X-ray fluxes above $5\times10^{-13}$~erg~s$^{-1}$~cm$^{-2}$ located within $0.5$~kpc of the Sun, matching the faintest and most distant observed XDINS up to now. To focus on the older population, we further exclude neutron stars in our simulations that are younger than $10^{5}$~yr, as the youngest XDINS with an independently estimated age from its kinematics (RX~J1856.5$-$3754 \cite{Tetzlaff2011}) is about $4\times10^{5}$ yr old.\\

We do not include additional observational biases, such as flux smearing for rotation-powered pulsars or detection only during outbursts for magnetically powered neutron stars, for two main reasons. First, modeling these biases for such a young population is complex, since most sources in \tablename~\ref{tab:young_ns} were discovered through targeted searches, making it difficult to account for them consistently. Second, including these effects has little impact on the results, as these young stars are typically bright, with spin periods and magnetic fields that have not yet decayed significantly. 
%Consequently, since we neglect several detection biases, our simulated sample likely overestimates the number of detectable sources. The resulting magnetar fraction should thus be regarded as a conservative estimate.\\

We perform 1000 simulations for each combination of \ac{CC-SN} rate, initial mean magnetic field of magnetars, and magnetar fraction at birth for the young population, and 100 simulations for the old population. For each simulated population, we determine the number of detectable magnetars, rotation-powered neutron stars, and XDINS according to the criteria described above. The results are shown in \figurename~\ref{fig:pop_sin_sim}, where orange dots, green stars, and blue dots with error bars denote the median and standard deviation of detected rotation-powered neutron stars (RPPs), magnetars, and XDINSs over the full set of simulations, respectively. Rows correspond to different \ac{CC-SN}e rates, while columns correspond to the four initial magnetic field strengths considered for magnetars, and compatible with the XDINSs (see above). Observed counts of magnetars, RPPs, and XDINSs are overlaid as solid green, orange, and blue lines, respectively, with an additional dashed green line including \ac{CCO}s. This choice is motivated by the possibility that some CCOs are young magnetars with weak external dipole fields but strong internal crustal fields \citep{Bogdanov2014, Popov2015}. Accordingly, we also estimate the magnetar fraction under the assumption that CCOs were born with magnetar-like fields.\\

Plausible values of the magnetar fraction for a given neutron-star class and \ac{CC-SN} rate correspond to cases in which the simulated number of detectable sources exceed the observed number of sources. In \figurename~\ref{fig:pop_sin_sim}, the vertically shaded green bands mark the ranges of magnetar fraction for which this condition is satisfied. Because the magnetar fraction must simultaneously reproduce the observed numbers of all neutron-star classes, the allowed values correspond to the overlap of the green shaded regions across the relevant panels (i.e., within each row and column of \figurename~\ref{fig:pop_sin_sim}). This overlapping region limited by the vertical dashed lines, therefore defines the magnetar fraction consistent with the joint constraints from young RPPs, magnetars, CCOs, and XDINS. We further test whether adopting a smaller standard deviation, $\sigma = 0.3$, for the initial magnetic-field distribution of magnetars significantly affects our results. We find that using this narrower distribution does not lead to substantial changes: the inferred magnetar fraction decreases slightly for weaker magnetic field strengths. For instance, for $B_{0,\rm mag} = 10^{14}$, it is reduced from 40–70\% to 40–60\%. Therefore, lower values of $\sigma$ result in a narrower range of magnetar fractions, and the value obtained for $\sigma = 0.5$ can be regarded as a conservative estimate.

\end{methods}

%ATT: Bibliography style
% Use this up until the time for final submission, then do the final compilation and download the bbl file, and comment out this line, put bibitems below separated for Article and Methods.

%\input{biblio_macros_2.tex}
%\bibliography{biblio.bib}

\begin{addendum}
 \item C.P.A. and N.R. are supported by the ERC via the Consolidator grant ``MAGNESIA'' (No. 817661), the ERC Proof of Concept "DeepSpacePULSE" (No. 101189496), and by the program Unidad de Excelencia Mar\'ia de Maeztu CEX2020-001058-M. M.R. is supported by the Dutch Research Council (NWO) via the grant CORTEX (NWA.1160.18.316) of the research programme NWA-ORC. V.G. is supported by a UKRI Future Leaders Fellowship (grant number MR/Y018257/1). We also acknowledge support from the Catalan grant SGR2021-01269 (PI: Graber/Rea) and the Spanish grant PID2023-153099NA-I00 (PI: Coti Zelati). C.P.A.’s work has been carried out within the framework of the doctoral program in Physics at the Universitat Autonoma de Barcelona. 
 The data production, processing, and analysis tools for this paper have been implemented and operated at the Port d’Informaci\'o Cient\'ifica (PIC) data center. PIC is maintained through a collaboration of the Institut de F\'isica d’Altes Energies (IFAE) and the Centro de Investigaciones Energ\'eticas, Medioambientales y Tecnol\'ogicas (Ciemat). We particularly thank Christian Neissner and Martin Børstad Eriksen for their support at PIC.
 We made use of the pulsar population synthesis code ML-Poppyns \cite{Ronchi2021, Graber2024, Pardo2025} funded by the European Research Council via the ERC Consolidator grant 'MAGNESIA' (No. 817661; PI: N. Rea), and publicly available at \url{https://ice-csic-astroexotic.github.io/code/ml_poppyns/}.
 The authors thank Samar Safi-Harb and Maria Kopsacheili for useful exchanges on supernova remnants age estimates, Clara Dehman for providing magnetic field evolution curves, Davide De Grandis and Francesco Coti Zelati for useful insights about magnetar outbursts and XDINS detectability, and Emilie Parent and Jos\'e Pons for useful discussions about the young pulsar populations. We thank Rui-Chong Hu and Bing Zhang for sharing their results on magnetar formation channels and for useful discussion.

  \item[Author Contributions]C.P.A. performed the population synthesis simulations, collected the observational data, wrote the methods section and produced figures and tables. N.R. conceived the project and wrote the main manuscript and discussion. M.R. helped with the population synthesis analysis and methods writing. V.G. contributed in work planning and population synthesis code. All authors have contributed to the discussion and interpretation of the results.

  \item[Competing Interests] The authors declare that they have no competing financial interests.
 \item[Data Availability]  Data products can be supplied by the authors on request.
  \item[Code Availability]  The code is available at \url{https://github.com/ice-csic-astroexotic/ML-Poppyns-Open}. Further code that supports this paper are available upon request to the authors.
   \item[Correspondence] Correspondence and requests for materials should be addressed to C.P.A, N.R., M.R. and V.G. (pardo@ice.csic.es, rea@ice.csic.es, ronchi@astron.nl, Vanessa.Graber@rhul.ac.uk)

\end{addendum}

\newpage
% ------ FIGURE --------------------
\begin{figure*}
\centering
\includegraphics[width=0.48\textwidth]{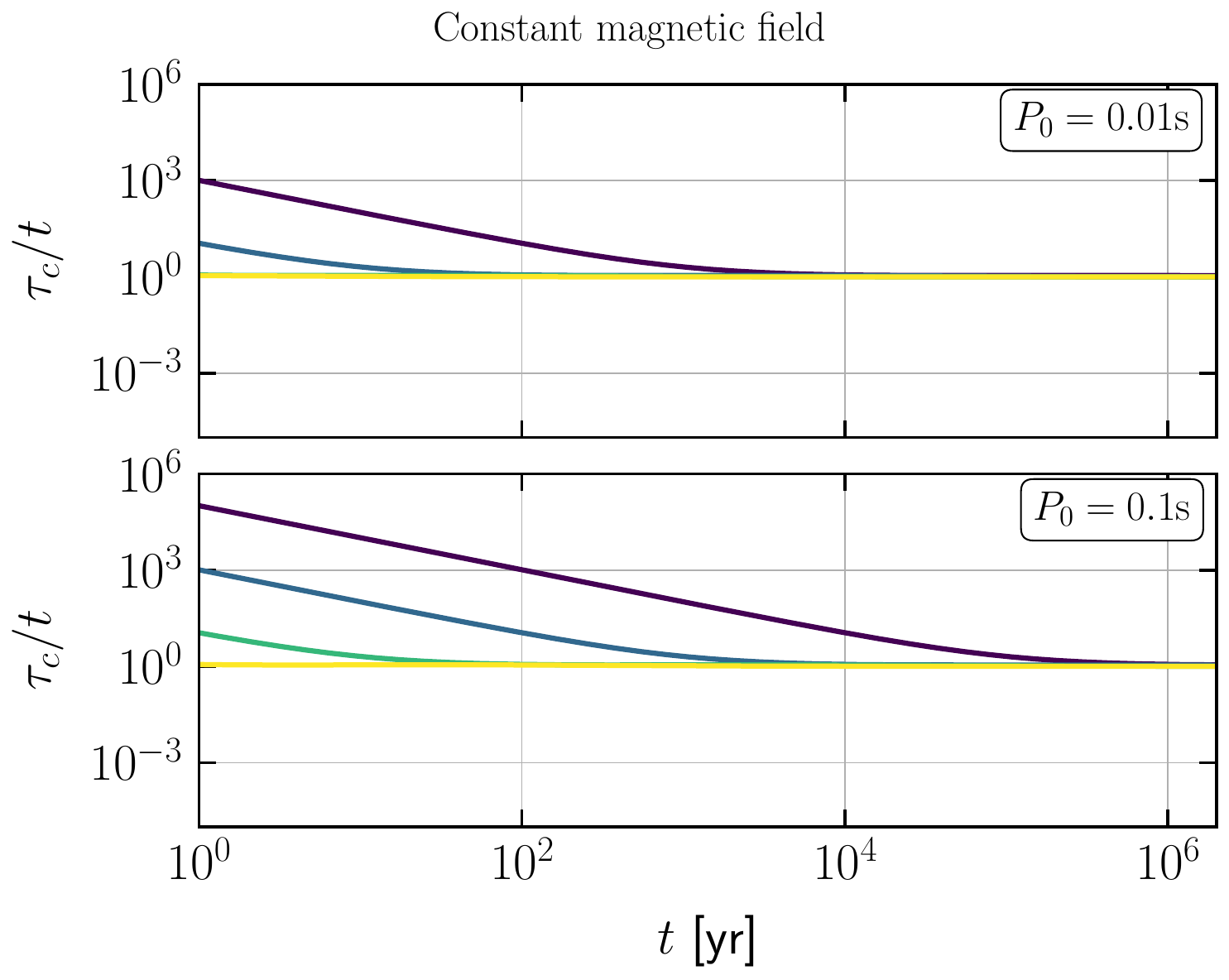}
\hfill
\includegraphics[width=0.48\textwidth]{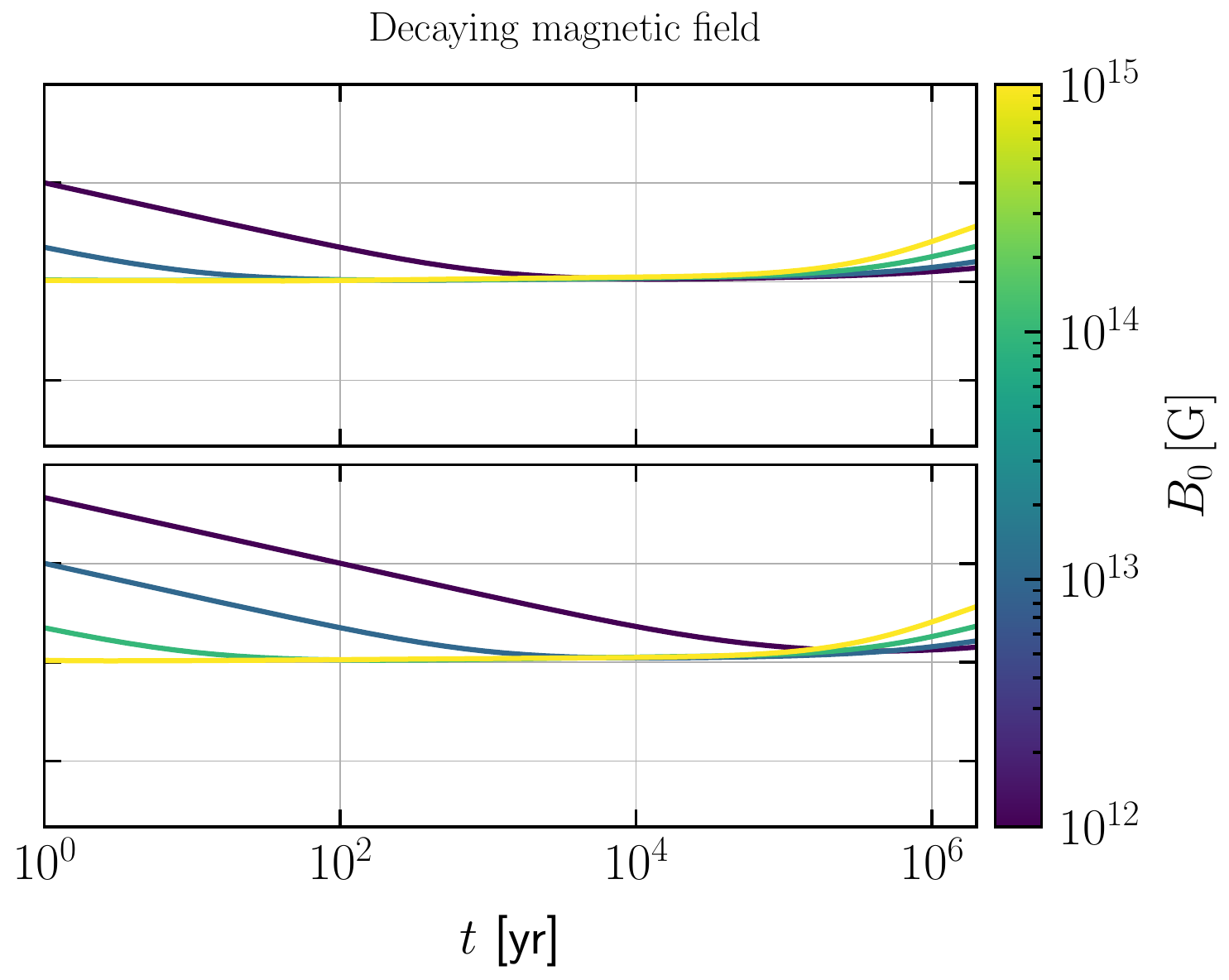}
\caption{
Ratio of characteristic age, $\tau_c$, vs.\ real age, $t$, for four different initial magnetic fields and two spin periods as a function of real age, $t$. The rows correspond to initial spin periods of 0.01 s (top) and 0.1 s (bottom). Colors indicate the initial magnetic field strengths. The left and right panels show cases of a constant and a decaying magnetic field, respectively.
}
\label{fig:charact_vs_real_age}
\end{figure*}

% ------------------------------------

% ------ FIGURE --------------------
\begin{figure}
\centering
    \includegraphics[width=0.7\linewidth]{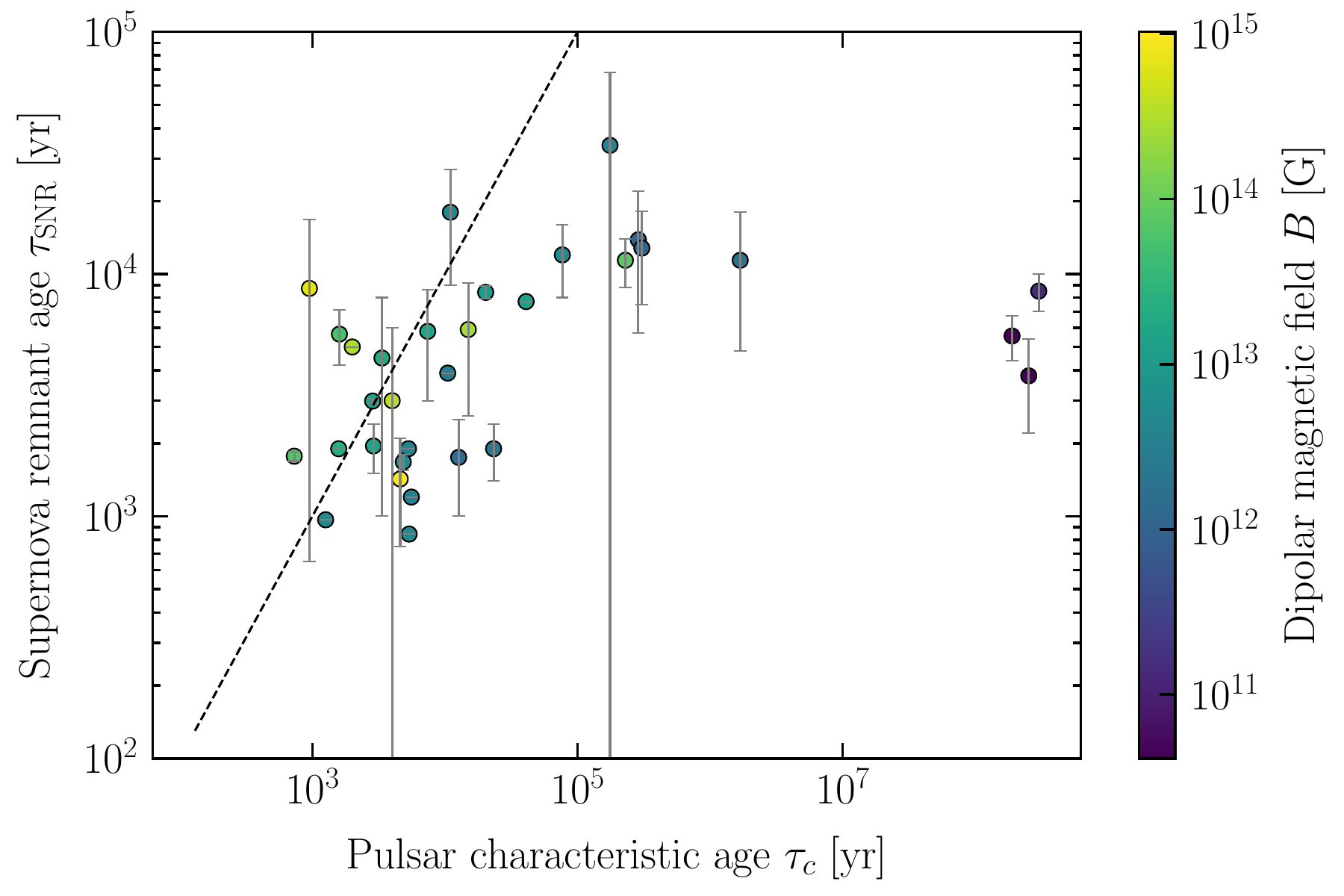}
    \caption{Neutron star characteristic age, $\tau_c$, versus the age inferred from the associated \ac{SNR}, $\tau_{\mathrm{SNR}}$. The color scale represents the dipolar magnetic field strength $B$ in Gauss. The dashed line indicates where $\tau_{\mathrm{SNR}} = \tau_c$. }

    \label{fig:characte_vs_real_b_dip}
\end{figure}
% ------------------------------------
%-------------------------------------------------------------------------------------
\begin{figure}[t!]
    \centering
    \includegraphics[width=0.7\linewidth]{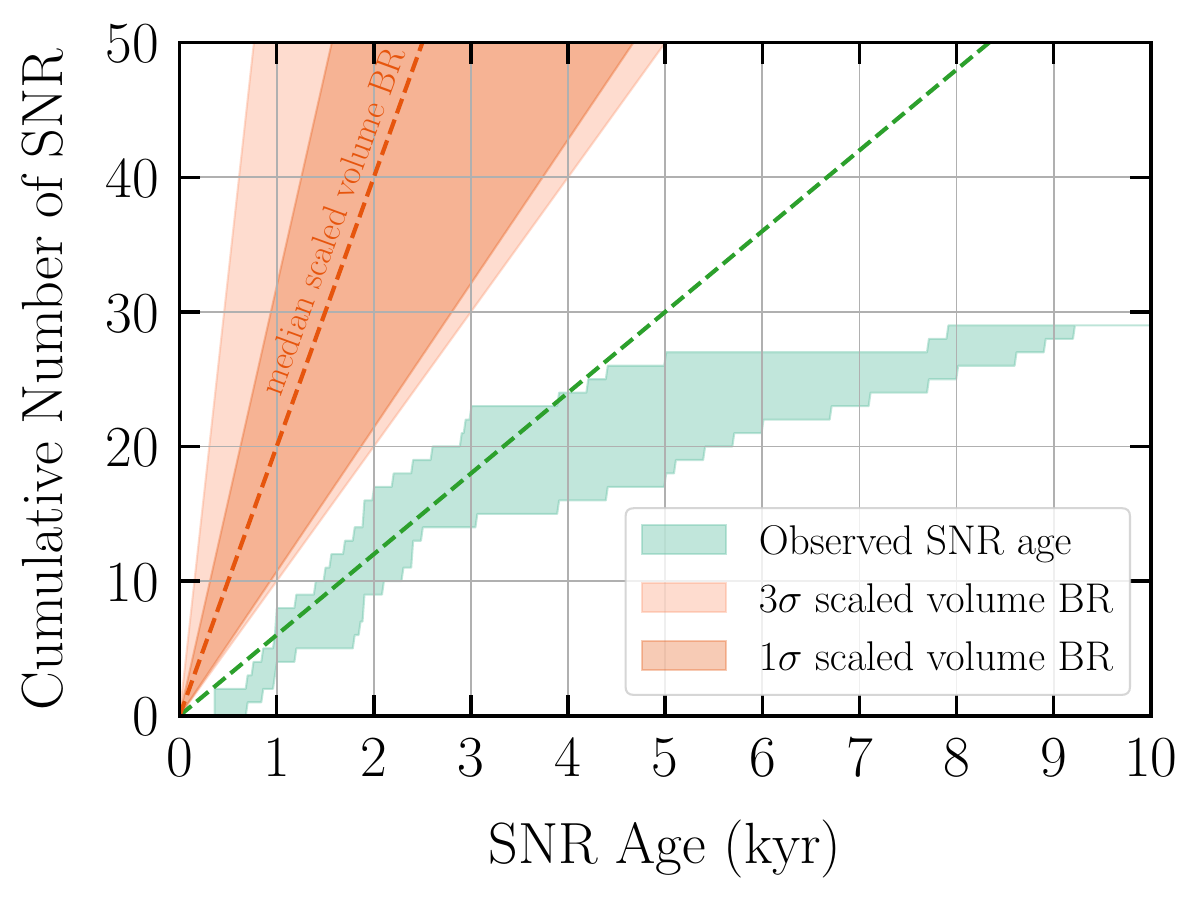}
    \caption{Cumulative age distribution of SNRs with neutron star associations. The green shaded area represents the cumulative distribution of observed \ac{SNR} ages, incorporating the lower and upper bounds derived from observational uncertainties. A green dashed line indicates a constant supernova birthrate of 0.6 events per century. The orange shaded region shows the volume-scaled birthrate estimated from a full evolutionary model of Galactic SNRs. The darker and lighter orange bands correspond to the $1\sigma$ and $3\sigma$ confidence intervals, respectively, while the orange dashed line marks the median of the estimated distribution obtained from 500K simulations. }
    \label{fig:age_cum}
\end{figure}
%-------------------------------------------------------------------------------------
%--------------------------------------------------------
\begin{figure*}
    \centering
    \includegraphics[width=1.2\linewidth]{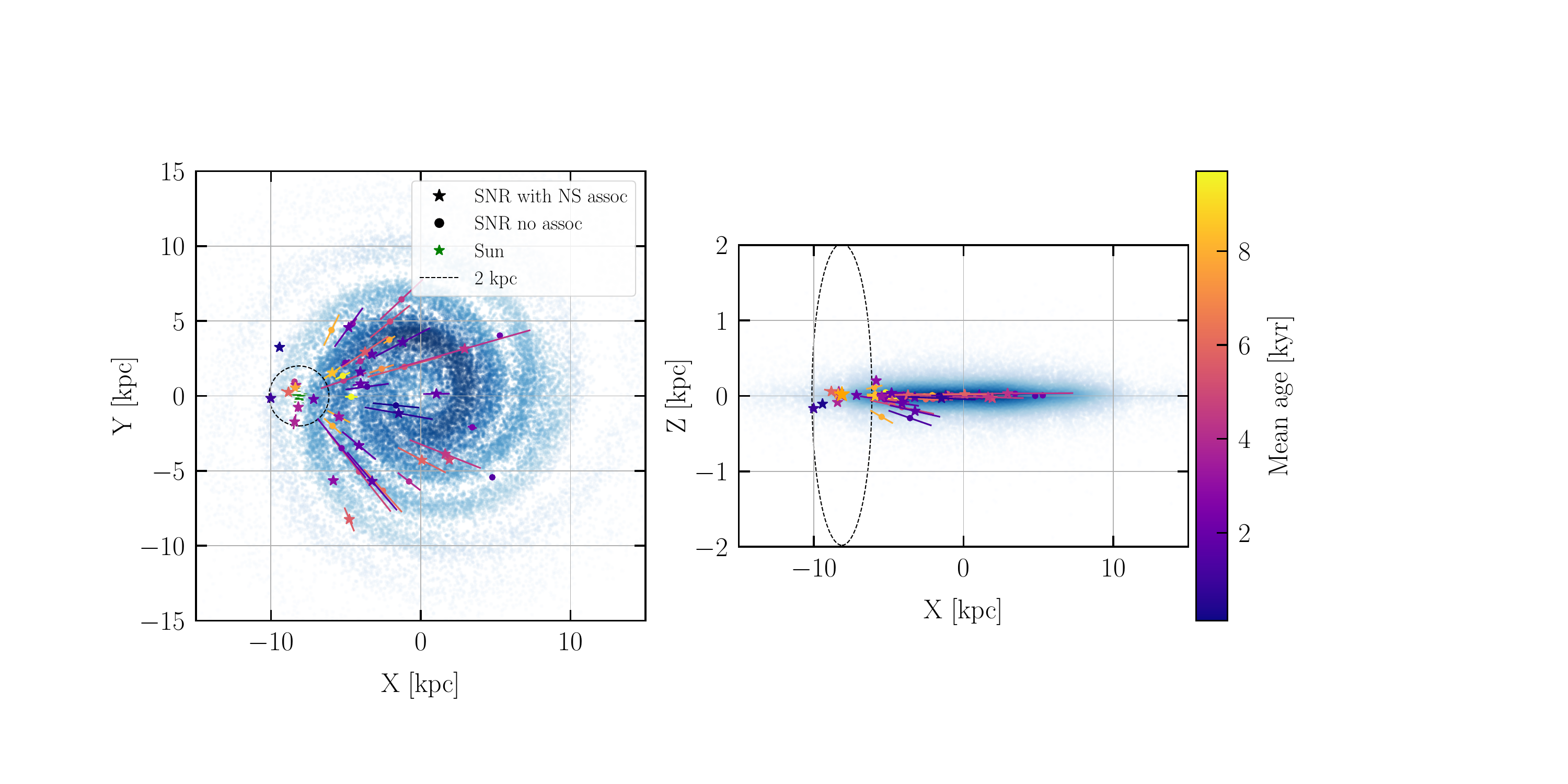}
    \caption{Spatial distribution of Galactic \ac{SNR}s overlaid on a simulated Galaxy shown in blue. Stars indicate \ac{SNR}s with associated neutron stars, while dots represent remnants without associated neutron stars. Solid lines denote the uncertainties in the estimated distances. Colors correspond to the mean age of each \ac{SNR}. The black dashed circle marks the 2 kpc radius around the Sun.}
    \label{fig:spatial_dist_snr}
\end{figure*}
%-------------------------------------------------------------------------------------
% ------  TABLE ------------------------------
\begin{table*}
\footnotesize
\setlength{\tabcolsep}{2pt}
\centering
\caption{Supernova remnants with neutron star associations younger than 10 kyrs within 2 kpc of the Sun}
\label{tab:snr_2kpc}
\begin{tabular}{l l l l l l l l l l l r}  
\hline\hline       
&SNR & NS & SNR Age & dist &$\tau_c$& Period & $B_{\rm dip}$ & $\dot{E}_{\rm rot}$ &  Type\\ 
&  & & [kyr] & [kpc] & [kyr] & [s] & $10^{14}\,\mathrm{G}$ & $[\mathrm{erg}\,\mathrm{s}^{-1}]$ &  \\
\hline  

1&G266.2-01.2 & CXOJ085201.4-461753 & 2.4–5.1 &   0.5–$1.0$\cite{Allen2015} & & & & &CCO  \\
2&G347.3-00.5 & 1WGAJ1713.4-3949 & 1.0–1.7 &   0.9–$1.1$\cite{Tsuji2016} & & & && CCO\\    
3& G184.6-05.8 (Crab) & J0534+2200 & 0.97 &  1.7–$2.1$\cite{Lin2023} & 1.26 & 0.033 & 0.05066 &	$7.9\times 10^{38}$	 &RPP \\
4& G260.4-03.4 & RXJ0822.0-4300 & 2.2–5.4 &  1.3–$2.2$\cite{Ranasinghe2022} & 254000 & 0.113 &0.00044 &	$4.5\times 10^{32}$& CCO \\
5 & G263.9-03.3 (Vela) & B0833-45 & 9.0–27.0 & 0.25–$0.3$\cite{Suzuki2021} & 11 &  0.089 & 0.04415 &$1.2\times 10^{37}$	 &RPP  \\
\hline
\multicolumn{9}{l}{\textbf{Dubious associations or uncertain estimates of age or distance}} \\
\hline
6&G106.3+02.7 & J2229+6114 & $>3.9$ & 0.7–$0.8$\cite{Kothes2006,Pope2024} & 10.5 &  0.052 & 0.027 &	$3.9\times 10^{37}$ & RPP \\
7&G114.3+00.3 & B2334+61 & 7.7 & 0.5–$0.7$\cite{Yar-Uyaniker2004} & 41 & 0.495 &0.131	& $1.1\times 10^{35}$ & RPP   \\
8&G160.9+02.6 & SGRJ0501+4516 & 2.6–9.2 & 0.3–$1.2$\cite{Ranasinghe2022} & 15 & 5.76 & 2.499 &	$2.2\times 10^{33}$ & Mag \\
\end{tabular}
The columns present the same information as in Table~\ref{tab:young_ns}, except that in this case the listed references correspond to the distance estimates.
\end{table*}
% ------------------------------------

% ---- TABLE -------------------------------
\begin{figure}
    \centering
    \includegraphics[width=0.6\linewidth]{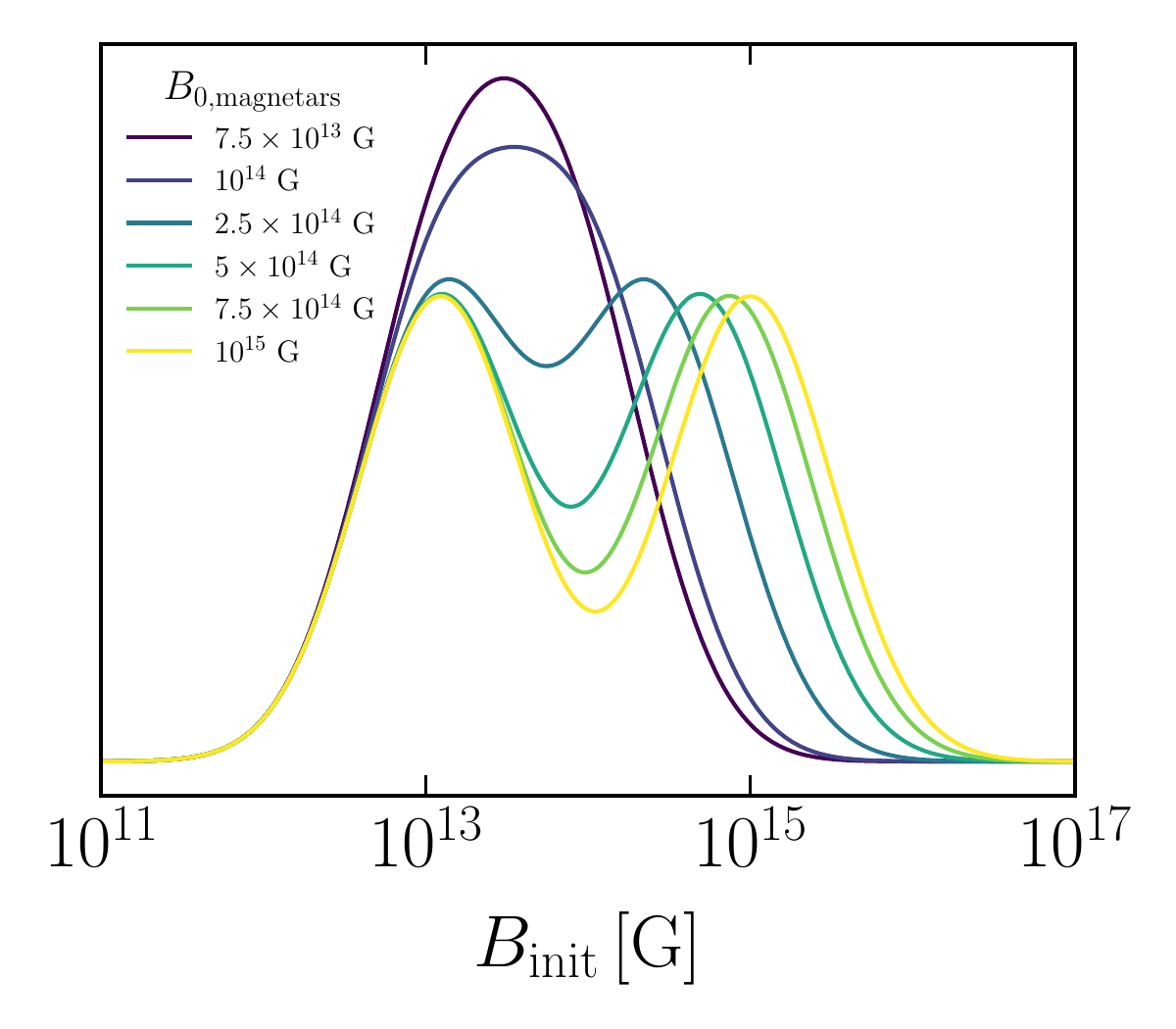}
   \caption{
    Magnetic-field distribution at birth modeled as a double log-normal. 
    The first component, representing the radio-pulsar population, uses 
    the parameters from Pardo-Araujo et al.\ (2025)~\cite{Pardo2025}. 
    The second component corresponds to magnetars and is shown for a range 
    of mean magnetic-field values, each modeled with a fixed width of 
    $\sigma = 0.5$. The relative normalization of the two components yields 
    a magnetar fraction of 50\% in this example.
    }
    \label{fig:init_mag}
\end{figure}

% --------- FIGURE -----------------------
\begin{figure*}
    \centering
    \includegraphics[width=1.1\linewidth]{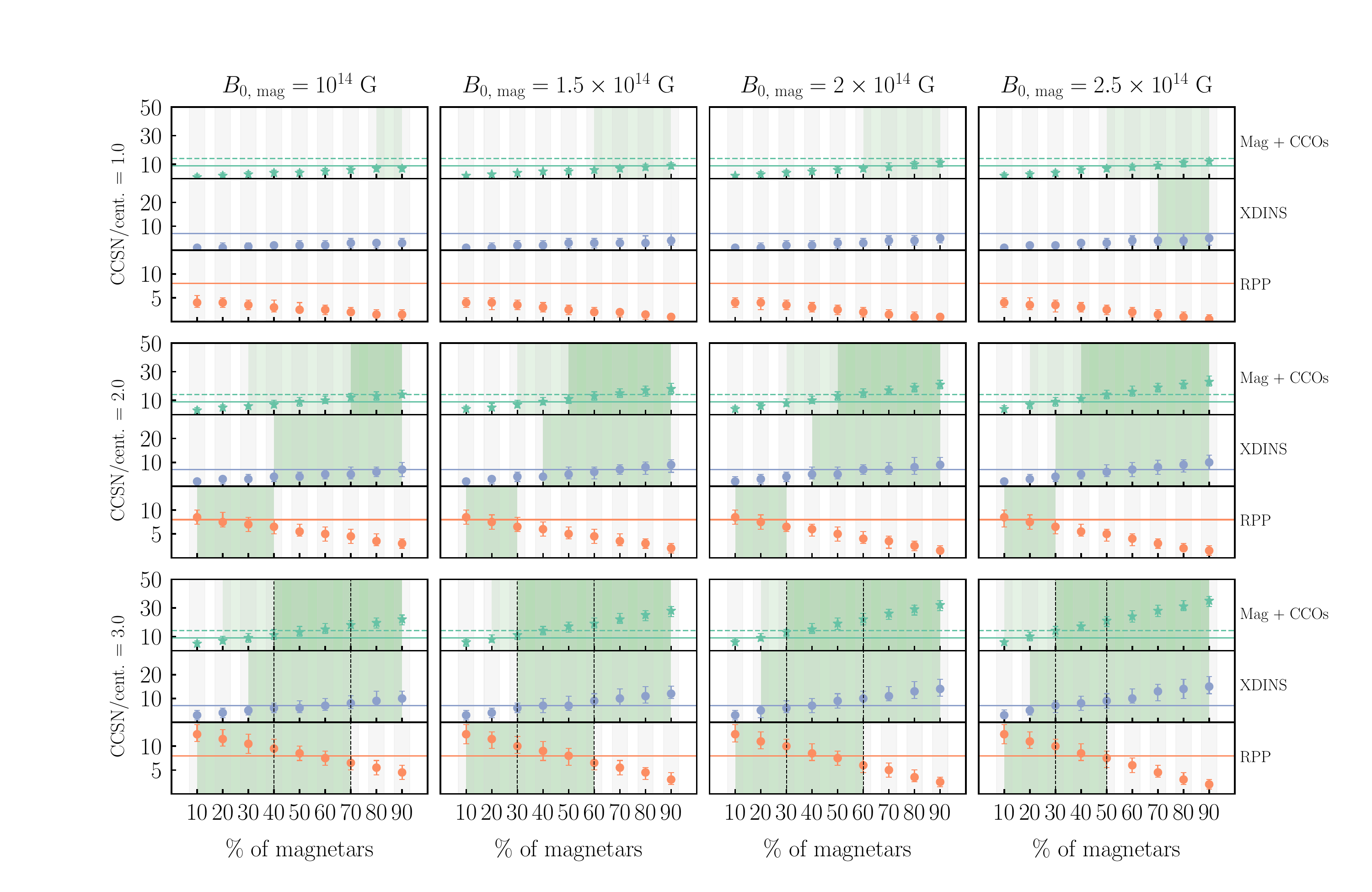}
    \caption{Simulated numbers of magnetars, rotation-powered neutron stars, and XDINS as a function of the magnetar birth fraction are shown for different mean initial magnetar magnetic fields and \ac{CC-SN} rates. Orange circles, green stars, and blue dots indicate the simulated counts of rotation-powered neutron stars, magnetars, and XDINS, respectively. Solid horizontal lines mark the corresponding observed numbers, with the dashed green line indicating the combined magnetar plus CCO count. Points and error bars represent the median and standard deviation of the simulations. Light-green shading marks magnetar birth fractions consistent with each class, and black dashed squares indicate the plausible range considering all classes.}\label{fig:pop_sin_sim}

\end{figure*}
%--------------------------------------------------------------------------------
\begin{figure}[t!]
    \centering
    \includegraphics[width=\linewidth]{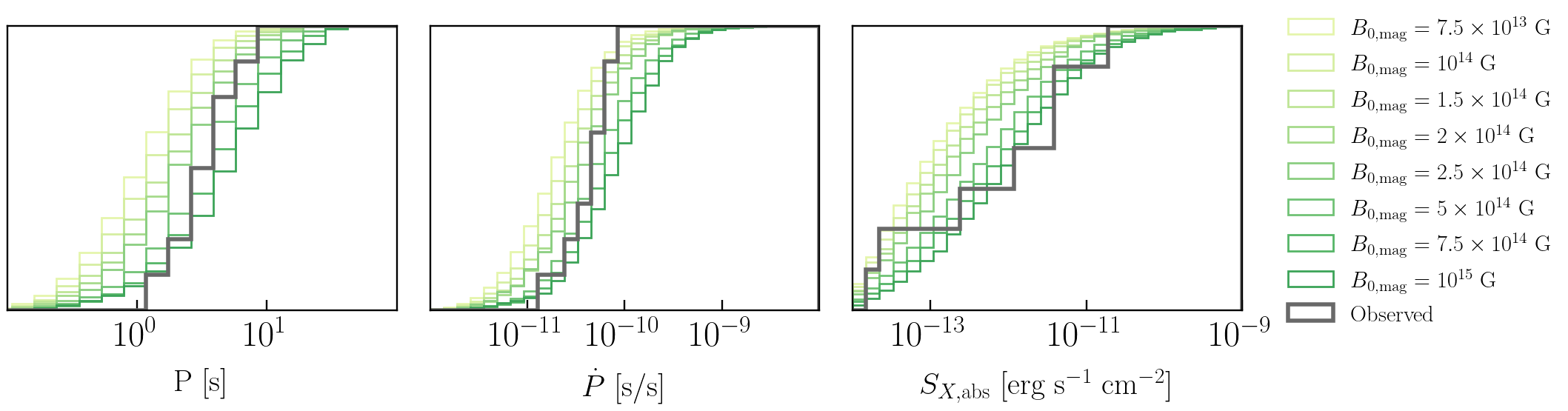}

    \vspace{0.3cm}

    \includegraphics[width=\linewidth]{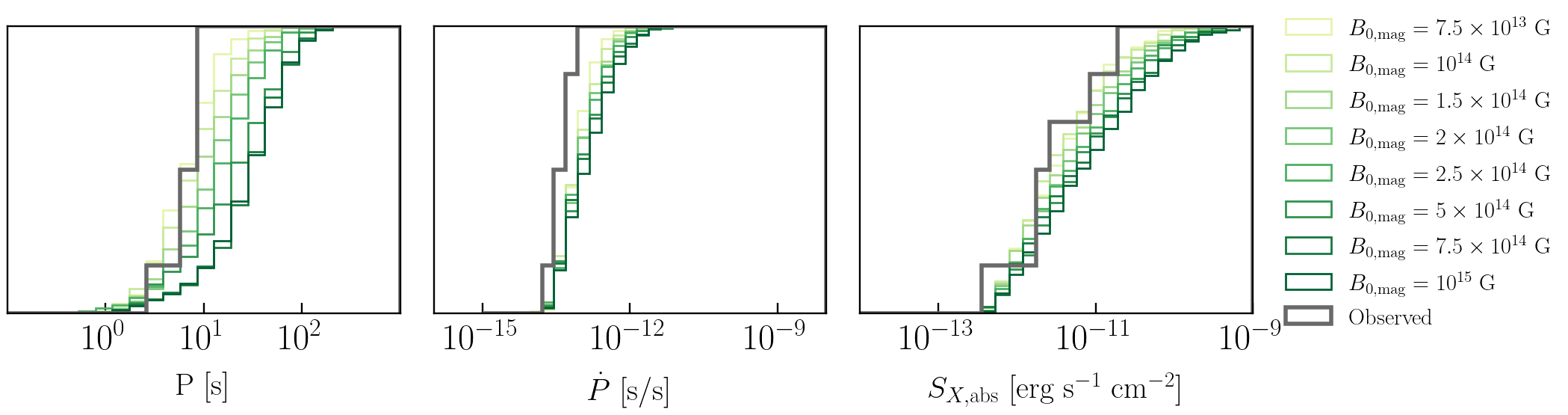}

    \caption{Cumulative distributions of the period ($P$), period derivative ($\dot{P}$), and absorbed X-ray flux ($S_{X,\rm abs}$) for observed and simulated neutron star populations. 
    \textbf{Top row:} magnetars. 
    \textbf{Bottom row:} X-ray dim isolated neutron stars (XDINS). 
    The green color gradient represents different initial magnetic fields for the magnetar component, while the solid black line shows the observed distributions.}
    \label{fig:cumulative_distributions}
\end{figure}

% ------------------------------------

% ------ TABLE --------------------
% ------ TABLE --------------------
\begin{table*}
\footnotesize
\setlength{\tabcolsep}{5pt}
\centering
\caption{Results of the two-sample 1D Kolmogorov--Smirnov test performed on the distributions of spin period, $P$, spin period derivative, $\dot{P}$, and X-ray absorbed flux $S_{\rm X, abs}$ for the simulated and observed magnetars. 
Shaded rows indicate cases where all three tests on the three distributions are valid at the 99\% confidence level ($p>0.01$).}
\label{tab:ks_magnetars}
\begin{tabular}{l ccc ccc}
\hline\hline
$B_{0,\rm mag}$ [G] &
KS($P$) & KS($\dot P$) & KS($S_{X,\rm abs}$) &
$p_P$ & $p_{\dot P}$ & $p_{S_{X,\rm abs}}$ \\
\hline
$7.5\times10^{13}$     & 0.627 & 0.513 & 0.380 & $1.423\!\times10^{-3}$ & $1.785\!\times10^{-2}$ & $1.504\!\times10^{-1}$ \\
$1.0\times10^{14}$     & 0.575 & 0.437 & 0.368 & $4.969\!\times10^{-3}$ & $6.636\!\times10^{-2}$ & $1.789\!\times10^{-1}$ \\
\rowcolor{gray!15}
$1.5\times10^{14}$     & 0.507 & 0.352 & 0.337 & $2.002\!\times10^{-2}$ & $2.169\!\times10^{-1}$ & $2.602\!\times10^{-1}$ \\
\rowcolor{gray!15}
$2.0\times10^{14}$     & 0.456 & 0.300 & 0.306 & $4.887\!\times10^{-2}$ & $3.919\!\times10^{-1}$ & $3.670\!\times10^{-1}$ \\
\rowcolor{gray!15}
$2.5\times10^{14}$     & 0.397 & 0.231 & 0.265 & $1.239\!\times10^{-1}$ & $7.151\!\times10^{-1}$ & $5.505\!\times10^{-1}$ \\
\rowcolor{gray!15}
$5.0\times10^{14}$     & 0.242 & 0.340 & 0.282 & $6.498\!\times10^{-1}$ & $2.500\!\times10^{-1}$ & $4.665\!\times10^{-1}$ \\
\rowcolor{gray!15}
$7.5\times10^{14}$     & 0.227 & 0.415 & 0.302 & $7.280\!\times10^{-1}$ & $9.323\!\times10^{-2}$ & $3.830\!\times10^{-1}$ \\
\rowcolor{gray!15}
$1.0\times10^{15}$     & 0.302 & 0.497 & 0.325 & $3.870\!\times10^{-1}$ & $2.468\!\times10^{-2}$ & $3.025\!\times10^{-1}$ \\
\hline
\end{tabular}
\end{table*}

% ------------------------------------
\begin{table*}
\footnotesize
\setlength{\tabcolsep}{5pt}
\centering
\caption{Results of the two-sample 1D Kolmogorov--Smirnov test performed on the distributions of spin period, $P$, spin period derivative, $\dot{P}$, and X-ray absorbed flux $S_{\rm X, abs}$ for the simulated and observed magnetars. 
Shaded rows indicate cases where all three tests on the three distributions are valid at the 99\% confidence level ($p>0.01$).}
\label{tab:ks_xdins}
\begin{tabular}{l ccc ccc}
\hline\hline
$B_{0,\rm mag}$ [G] &
KS($P$) & KS($\dot P$) & $KS(S_{X,\rm abs})$ &
$p_{P}$ & $p_{\dot P}$ & $p_{S_{X,\rm abs}}$ \\
\hline
\rowcolor{gray!15}
$7.5\times10^{13}$  & 0.316 & 0.495 & 0.245 & $5.166\times10^{-1}$ & $8.066\times10^{-2}$ & $8.075\times10^{-1}$ \\
\rowcolor{gray!15}
$1.0\times10^{14}$  & 0.407 & 0.477 & 0.279 & $2.094\times10^{-1}$ & $9.042\times10^{-2}$ & $6.507\times10^{-1}$ \\
\rowcolor{gray!15}
$1.5\times10^{14}$  & 0.503 & 0.516 & 0.228 & $6.444\times10^{-2}$ & $5.391\times10^{-2}$ & $8.541\times10^{-1}$ \\
\rowcolor{gray!15}
$2.0\times10^{14}$  & 0.611 & 0.527 & 0.271 & $1.140\times10^{-2}$ & $4.587\times10^{-2}$ & $6.841\times10^{-1}$ \\
\hline
$2.5\times10^{14}$  & 0.661 & 0.531 & 0.278 & $4.931\times10^{-3}$ & $4.559\times10^{-2}$ & $6.609\times10^{-1}$ \\
$5.0\times10^{14}$  & 0.730 & 0.584 & 0.316 & $1.360\times10^{-3}$ & $2.265\times10^{-2}$ & $5.109\times10^{-1}$ \\
$7.5\times10^{14}$  & 0.849 & 0.530 & 0.306 & $3.873\times10^{-5}$ & $4.806\times10^{-2}$ & $5.488\times10^{-1}$ \\
$1.0\times10^{15}$  & 0.853 & 0.602 & 0.357 & $2.549\times10^{-5}$ & $1.405\times10^{-2}$ & $3.531\times10^{-1}$ \\
\hline
\end{tabular}
\end{table*}

% --------------------------------------
% --------------------------------------
% --------------------------------------

% --------------------------------------

\end{document}